\documentclass[12pt]{article}
\usepackage{amssymb}
\usepackage{amsmath}
\usepackage{amstext}
\usepackage{graphicx,epsfig}
\usepackage{epsfig}
\usepackage{color}
\usepackage{ulem}
\usepackage{parskip}
\usepackage{cite}
\usepackage{tikz}
\usepackage{mathtools}
\usepackage[utf8]{inputenc}
\usepackage{relsize}

\usetikzlibrary{positioning, trees,decorations, decorations.markings, decorations.pathmorphing, arrows, shapes.geometric, snakes}

\allowdisplaybreaks

\newcommand{\Comment}[1]{{}}
\definecolor{MyDarkBlue}{rgb}{0.15,0.15,0.45}
\usepackage[linktocpage=true]{hyperref}
\hypersetup{
colorlinks=true,
citecolor=MyDarkBlue,
linkcolor=MyDarkBlue,
urlcolor=MyDarkBlue,
pdfauthor={Kurt Hinterbichler},
pdftitle={},
pdfsubject={hep-th}
}

\newcommand{\be}{\begin{equation}}
\newcommand{\ee}{\end{equation}}
\newcommand{\bal}{\begin{align}}
\newcommand{\eal}{\end{align}}
\newcommand{\bals}{\begin{align*}}
\newcommand{\eals}{\end{align*}}
\newcommand{\bea}{\begin{eqnarray}}
\newcommand{\eea}{\end{eqnarray}}
\newcommand{\beas}{\begin{eqnarray*}}
\newcommand{\eeas}{\end{eqnarray*}}
\newcommand{\nn}{\nonumber}

\def\({\left(}
\def\){\right)}

\newcommand{\ccdot}{\! \cdot \!}

\numberwithin{equation}{section}

\begin{document}

% Define styles for the different kind of edges in a Feynman diagram
\tikzset{
xparticle/.style={decorate , dashed, draw=black},
  mgraviton/.style={decorate, draw=black,
    decoration={coil,amplitude=4.5pt, segment length=7pt}}
}

\begin{center}
{\LARGE \bf{Constraints on a Gravitational}}\\
{\LARGE \bf{Higgs Mechanism\\ \vspace{.2cm} }}
\end{center}
\vspace{2truecm}
\thispagestyle{empty}
\centerline{{\large James Bonifacio,${}^{\rm a,}$\footnote{\href{mailto:james.bonifacio@case.edu}{\texttt{james.bonifacio@case.edu}}} Kurt Hinterbichler,${}^{\rm a,}$\footnote{\href{mailto:kurt.hinterbichler@case.edu}{\texttt{kurt.hinterbichler@case.edu}}} and Rachel A. Rosen${}^{\rm b,}$\footnote{\href{mailto:rar2172@columbia.edu}{\texttt{rar2172@columbia.edu}}}}}
\vspace{.5cm}
 
\centerline{{\it ${}^{\rm a}$CERCA, Department of Physics,}}
 \centerline{{\it Case Western Reserve University, 10900 Euclid Ave, Cleveland, OH 44106}} 
 \vspace{.25cm}
 
 \centerline{{\it ${}^{\rm b}$Center for Theoretical Physics, Department of Physics,}}
 \centerline{{\it Columbia University, New York, NY 10027}} 
 \vspace{.25cm}

 \vspace{.8cm}

\begin{abstract}
We show that it is impossible to improve the high-energy behavior of the tree-level four-point amplitude of a massive spin-2 particle by including the exchange of any number of scalars and vectors in four spacetime dimensions. This constrains possible weakly coupled ultraviolet extensions of massive gravity, ruling out gravitational analogues of the Higgs mechanism based on particles with spins less than two. Any tree-level ultraviolet extension that is Lorentz invariant and unitary must involve additional massive particles with spins greater than or equal to two, as in Kaluza-Klein theories and string theory. 

\end{abstract}

\newpage

\setcounter{tocdepth}{3}
\tableofcontents

\section{Introduction}
\parskip=5pt
\normalsize

The Higgs mechanism is a central feature of the standard model, the theory of superconductivity, and countless other more speculative scenarios.  The mechanism is often conceptualized in terms of spontaneous symmetry breaking: a gauge symmetry is broken by the vacuum expectation value of some scalar Higgs field, and the massless gauge fields ``eat" some components of the Higgs field to become massive, leaving behind physical scalars.   

Looking only at the S-matrix, we may think about the Higgs mechanism differently: it is a method of raising the ultraviolet (UV) strong coupling scale of an effective theory of self-interacting massive spin-1 particles by adding weakly coupled scalars to the theory.  For example, in the low-energy effective theory of $W^{\pm}$ and $Z^0$ massive vector bosons, the four-point amplitude of the longitudinal modes grows at high energies as $\sim E^2/v^2$, and violates perturbative unitarity when the center-of-mass energy $E$ becomes of order $v = 246 \, \rm{GeV}$. If this unitarity violation is to be cured while remaining weakly coupled, then another particle must enter before a scale of order $v$ and contribute to the tree amplitude in such a way as to cancel the bad high-energy growth.  The physical Higgs scalar is the simplest particle that accomplishes this cancellation, leading to an amplitude which does not grow with energy, and thus raising the strong coupling scale all the way to infinity.

A natural question is whether a similar mechanism exists for the gravitational field, i.e., for a spin-2 particle.  From the symmetry breaking point of view, this would be a mechanism in which a lower-spin Higgs field gets a vacuum expectation value which breaks the diffeomorphism symmetry of the massless graviton.  The graviton would then eat some of the Higgs field, becoming a massive graviton and leaving some other lower-spin physical fields left over.  Given that the global symmetry which is gauged to diffeomorphism symmetry is Poincar\'e symmetry, one might expect this gravitational Higgs mechanism to spontaneously break Poincar\'e symmetry.  Indeed, the ghost condensate can be understood along these lines \cite{ArkaniHamed:2003uy}.  

However, despite this intuition, we would like to know if there is a fully Poincar\'e-invariant gravitational Higgs mechanism. This question is an old one, and there are many previous proposals and studies, see for example \cite{Percacci:1990wy,Kakushadze:2000zn,Chamseddine:2003ft,Bandos:2003tm,Kirsch:2005st,Leclerc:2005qc,tHooft:2007rwo,Kakushadze:2007hf,Wever:2009laa,Pirinccioglu:2009bc,Chamseddine:2010ub,Oda:2010wn,Oda:2010gn,Berezhiani:2010xy,Iglesias:2011it, Blas:2014ira, Caron-Huot:2016icg, Torabian:2017bqu}. In terms of the S-matrix, the question is whether there is a method of raising the UV strong coupling scale of an effective field theory of self-interacting massive spin-2 bosons while remaining weakly coupled.   In analogy to the spin-1 case, we might expect that this can be done by adding lower-spin massive particles to the theory.

The goal of this paper is to determine in complete generality whether it is possible to introduce additional particles with spins less than two into the effective field theory of a single massive spin-2 particle so as to improve the high-energy behavior of the tree amplitudes and thus raise the strong coupling scale of the low-energy theory.  We know that the four-point tree-level amplitude in any effective theory of a massive spin-2 particle scales at least as badly as $\sim E^6$ at high energies~\cite{Bonifacio:2018vzv}, corresponding to a strong coupling scale of $\Lambda_3 = \left( m^2 M_p \right)^{1/3}$.  We will thus be asking whether this high-energy behavior can be softened at all without sacrificing Lorentz invariance or unitarity.  This is a weaker requirement than asking for a full UV completion, for which the amplitude would be bounded at high energies, so we will say that we are looking for a weakly-coupled UV {\it extension}.  

\begin{figure}[!ht]
\begin{center}
\epsfig{file=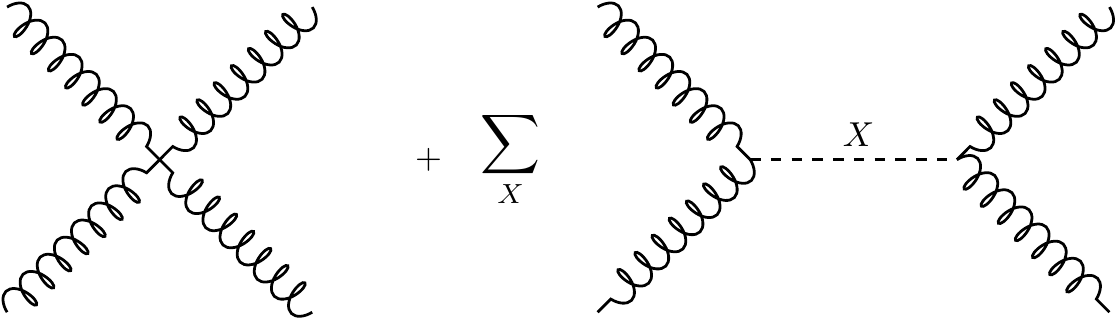,scale=1}
\caption{\small Schematic depiction of the four-point tree amplitude for a massive spin-2 particle. The sum is over exchanged states, $X$, which we allow to include the massive spin-2 particle, a graviton, and any number of scalars and vectors.}
\label{fig:amp}
\end{center}
\end{figure}

Our approach is to study in a model-independent way the high-energy behavior of the tree-level four-point amplitude of a massive spin-2 particle, allowing for the exchange of various other particles, as depicted in Fig.~\ref{fig:amp}.\footnote{A similar calculation, but restricting to operators with dimensions $\leq 4$ and bounded amplitudes, was presented in Ref.~\cite{Christensen:2014wra}.} Our conclusion will be that there is no way to improve the high-energy behavior of the four-point amplitude by exchanging any finite number of spin-0 and spin-1 particles in four spacetime dimensions.\footnote{An argument against UV extending massive gravity up to $M_p$ with a Higgs mechanism is given in Ref.~\cite{Arkani-Hamed:2017jhn}, namely that at high energies the massive graviton's longitudinal mode does not couple to its tensor modes with the interactions dictated by the equivalence principle. However, as pointed out in Ref.~\cite{deRham:2018qqo}, the equivalence principle constraints do not apply straightforwardly in the massless limit, since departures from masslessness can be important due to factors of the inverse mass occurring in interactions.}  This remains true even if we include a massless spin-2 particle in the spectrum.

Our results apply to massive gravity and bigravity theories.  Finding UV completions of these theories remains an important open problem in the field.  Important clues about possible UV completions of massive gravity or bigravity come from positivity and causality constraints~\cite{Adams:2006sv,Cheung:2016yqr,Camanho:2016opx,Bellazzini:2016xrt,Bellazzini:2017fep, deRham:2017zjm, deRham:2017xox,Hinterbichler:2017qyt,Bonifacio:2017nnt,deRham:2018qqo, Afkhami-Jeddi:2018apj}. 
These constraints pertain to full weakly-coupled UV completions that, according to conventional wisdom, would require introducing infinitely many new particles with arbitrarily high spins.  See, for example, Ref.~\cite{deRham:2018qqo} for an explicit argument using the Froissart bound.  Our results are completely consistent with these arguments.  Ghost-free theories of massive gravity \cite{deRham:2010ik,deRham:2010kj,Hassan:2011hr,Hinterbichler:2011tt,deRham:2014zqa}, bigravity \cite{Hassan:2011zd}, and multigravity \cite{Hinterbichler:2012cn} have improved high-energy behavior compared to generic ghostly theories of massive spin-2 particles, becoming strongly coupled around the scale $\Lambda_3 = \left( m^2 M_p \right)^{1/3}$~\cite{ArkaniHamed:2002sp,Schwartz:2003vj}. 
Our conclusions imply that for massive gravity and bigravity, any tree-level UV extension must contain additional massive particles with spins greater than or equal to two. 
This conclusion is consistent with the arguments concerning full weakly coupled UV completions, but is a stronger statement as it concerns a UV extension and as we need no assumptions about the asymptotic behavior of the full S-matrix.  Explicit examples of theories with massive spin-2 particles and improved UV behavior are Kaluza-Klein theories, string theory and large-$N$ QCD.  In each of these examples, an infinite tower of massive higher-spin particles (i.e., $s \geq 2$)  appear in the theory with masses that are parametrically close to that of the spin-2 particle.

{\bf Conventions:}
We work in four spacetime dimensions and use the mostly plus metric signature. The four-dimensional epsilon symbol is defined with $\varepsilon_{0123} =1$. Conventions on kinematics and polarizations are detailed in Appendix \ref{app:details}.

\section{Massive gravity coupled to scalars and vectors}
\label{sec:part1}

We begin in this section by showing that the best high-energy behavior of the four-point amplitude in a theory of massive gravity coupled to scalars and vectors is $\sim E^6$.  In using the words ``massive gravity,'' we are assuming that the massive spin-2 interactions have the form of the Einstein-Hilbert kinetic term plus a potential. This is not the most general case, but it will serve as a good warm-up for the general argument presented in Section \ref{sec:part2} since we will be able to show intermediate steps of the calculation and more easily visualize what is going on in terms of a Lagrangian.

\subsection{Interactions}

We consider a massive gravity Lagrangian given by,
\be \label{eq:mgL}
\mathcal{L}_{\rm mg} = \frac{M_p^2}{2} \sqrt{-g}\left( R -\frac{1}{4}m^2 V(g,h) \right),
\ee
where the metric is $g_{\mu \nu}=\eta_{\mu \nu}+ h_{\mu \nu}$ and $m$ is the graviton mass. The potential $V(g,h)$ can be expanded as
\begin{align}
V(g,h) =& \langle h^2 \rangle - \langle h \rangle^2+c_1 \langle h^3 \rangle +c_2 \langle h^2 \rangle \langle h \rangle+c_3 \langle h \rangle^3 \\
& +d_1 \langle h^4 \rangle+d_2 \langle h^3 \rangle\langle h \rangle+d_3 \langle h^2 \rangle^2+d_4 \langle h^2 \rangle\langle h \rangle^2+d_5 \langle h \rangle^4+ \ldots,
\end{align}
where angled brackets denote traces of matrix products with indices raised using $g^{\mu \nu}$, e.g. $\langle h^2 \rangle = g^{\mu \nu} g^{\lambda \rho} h_{\mu \lambda} h_{\nu \rho}$.   The quadratic term is fixed to the Fierz-Pauli form.  The canonically normalized massive spin-2 field is given by 
\be \hat{h}_{\mu \nu} = 2 h_{\mu \nu}/M_p\,.\ee
  All the terms up to fourth order in $h_{\mu\nu}$ are shown, with arbitrary coefficients in front of each.   Only the terms proportional to $c_1$, $c_2$, $d_1$, and $d_3$ contribute to the four-point scattering amplitude. 

We now add to the massive gravity Lagrangian a collection of scalar fields $\phi_j$ and vector fields $A_{j,\mu}$.  We consider both massive and massless scalars with masses $m_{\phi_j}$ and massive vectors with masses $m_{A_j}$.  The only new graphs that contribute to the tree-level four-point amplitude of massive gravitons are those that exchange one of the new particles.  These involve cubic interactions of the form $\hat{h}^2 \phi_j$ and $\hat{h}^2 A_j$, which can be both parity even and parity odd since we do not assume that parity is conserved. The most general interactions of this form are given by
\begin{align}
 \mathcal{L}_{\hat{h} \hat{h} \phi_j}  &=\frac{m^2}{2 M_p}  \sum_{l \geq 0}\Big( c_{1,l,j} \hat{h}_{\mu \nu} \hat{h}^{\mu \nu}+{c_{2,l,j}}m^{-2} \partial_{\lambda} \hat{h}_{\mu \nu} \partial^{\nu} \hat{h}^{\mu \lambda} + c_{3,l,j} m^{-4}\partial_{\lambda} \partial_{\rho} \hat{h}_{\mu \nu} \partial^{\mu} \partial^{\nu} \hat{h}^{\lambda \rho}  \nn \\
&+\tilde{c}_{1,l,j} m^{-2}\varepsilon^{\mu \nu \lambda \rho} \partial_{\mu} \hat{h}_{\lambda \sigma} \partial_{\nu} \hat{h}_{\rho}{}^{\sigma}+\tilde{c}_{2,l,j} m^{-4}\varepsilon^{\mu \nu \lambda \rho} \partial_{\mu} \partial^{\sigma} \hat{h}_{\lambda \gamma} \partial_{\nu}\partial^{\gamma} \hat{h}_{\rho \sigma}\Big) m^{-2l} \Box^l \phi_j,  \label{eq:hhphi} \\
\mathcal{L}_{\hat{h} \hat{h} A_j}&= \frac{m_{A_j}m}{M_p} \sum_{l \geq 0}\Big( d_{1,l,j}m^{-1}\hat{h}_{\mu \nu} \partial^{\mu} \hat{h}^{\nu \lambda}+d_{2,l,j}m^{-3} \partial_{\rho} \hat{h}_{\mu \nu} \partial^{\mu} \partial^{\nu} \hat{h}^{\rho \lambda} \nn \\
& + \tilde{d}_{1,l,j} m^{-1}\varepsilon^{\mu \nu \rho \lambda} \partial_{\mu} \hat{h}_{\nu \sigma} \hat{h}_{\rho}{}^{\sigma}+ \tilde{d}_{2,l,j}m^{-3} \varepsilon^{\mu \nu \rho \sigma} \partial_{\mu} \partial_{\gamma} \hat{h}_{\rho}{}^{\lambda} \partial_{\nu} \hat{h}_{\sigma}{}^{\gamma}\Big) m^{-2l} \Box^l A_{j, \lambda} \nn \\
&+ \frac{m_{A_j}}{2M_p} \sum_{l \geq 0}\Big(  d_{3,l,j} \hat{h}_{\mu \nu} \hat{h}^{\mu \nu}+d_{4,l,j} m^{-2}\partial_{\lambda} \hat{h}_{\mu \nu} \partial^{\nu} \hat{h}^{\mu \lambda} + d_{5,l,j} m^{-4}\partial_{\lambda} \partial_{\rho} \hat{h}_{\mu \nu} \partial^{\mu} \partial^{\nu} \hat{h}^{\lambda \rho} \nn \\
& + \tilde{d}_{3,l,j} m^{-2}\varepsilon^{\mu \nu \lambda \rho} \partial_{\mu} \hat{h}_{\lambda \sigma} \partial_{\nu} \hat{h}_{\rho}{}^{\sigma}+\tilde{d}_{4,l,j} m^{-4}\varepsilon^{\mu \nu \lambda \rho} \partial_{\mu} \partial^{\sigma} \hat{h}_{\lambda \gamma} \partial_{\nu}\partial^{\gamma} \hat{h}_{\rho \sigma} \Big) m^{-2l}\Box^l \partial^{\lambda} A_{j, \lambda}, \label{eq:hhA}
\end{align}
where $c_{i,l,j}$, $\tilde{c}_{i,l,j}$, $d_{i,l,j}$, and $\tilde{d}_{i,l,j}$ are real dimensionless coupling constants and the factors of $m_{A_j}$, $M_P$ extracted out front are to simplify later expressions. To obtain these interactions we modified the procedure for finding all on-shell cubic vertices, described in Appendix~\ref{app:details}, to allow the particle of lowest spin to be off shell. This amounts to  ignoring terms involving $\hat{h}_{\mu}{}^{\mu}$ and $\partial^{\mu} \hat{h}_{\mu \nu}$, and any terms that can be brought to this form by integration by parts, since these do not contribute when the massive spin-2 particle is an external leg. However, since the particles on the internal leg are off shell, we include interactions containing $\partial^{\lambda} A_{j,\lambda}$ and powers of $\Box$ acting on the lower-spin fields, which are equivalent to higher-order contact terms under a field redefinition and may contribute to the massive spin-2 four-point amplitude. 

The total Lagrangian we consider is thus
\be \label{eq:Ltot}
\mathcal{L} = \mathcal{L}_{\rm mg}+\sum_j \left( -\frac{1}{2} (\partial \phi_j)^2-\frac{1}{2} m_{\phi_j}^2\phi_j^2 + \mathcal{L}_{\hat{h} \hat{h} \phi_j}-\frac{1}{4}F^j_{\mu \nu}F_j^{\mu\nu}-\frac{1}{2}m_{A_j}^2 A_j^2+ \mathcal{L}_{\hat{h} \hat{h} A_j} \right)+ \ldots,
\ee
where $F^j_{\mu \nu} \equiv \partial_{\mu}A^j_{\nu} -\partial_{\nu} A^j_{\mu}$ and the terms not shown do not contribute to the four-point amplitude with external massive spin-2 particles. 
We assume that the total number of derivatives in the interactions is bounded above by $2N$ for some integer $N>1$, so the index $l$ in Eqs. \eqref{eq:hhphi} and \eqref{eq:hhA} has a finite range. This means that we do not consider the possibility of having infinitely many derivatives that resum into a function with soft high-energy behavior. 

\subsection{Amplitudes}
We now calculate the four graviton tree amplitude from this Lagrangian. For this calculation we use helicity polarizations and work in the center-of-mass frame (kinematic details and conventions are reviewed in Appendix~\ref{app:details}). We denote this amplitude by $\mathcal{A}_{h_1 h_2 h_3 h_4}$, where $h_j \in \{0,\pm1,\pm2\}$ denotes the helicity of particle $j$. Our aim is to fix the coupling constants so that the tree amplitudes have the best possible high-energy behavior for fixed-angle scattering. 

Consider the amplitudes with $h_1=h_3$ and $h_2=h_4$. In massive gravity these amplitudes grow with energy at worst like
\be
\mathcal{A}_{h_1 h_2 h_1 h_2}  \sim E^{10-2\left(|h_1|+|h_2|\right)}.
\ee
The scalar interactions with $2n$ derivatives or vector interactions with $2n-1$ derivatives produce exchange amplitudes that generically grow with energy like 
\be \label{eq:exchangepower}
\mathcal{A}_{h_1 h_2 h_1 h_2}  \sim E^{4n+6-2\left(|h_1|+|h_2|\right)}.
\ee
By comparing these, we see that the leading amplitudes produced by the scalar and vector interactions with more than two derivatives must cancel between themselves, otherwise the high-energy behavior would be as bad as or worse than in massive gravity. As we will see, this condition forces these higher-derivative interactions to vanish.

By an explicit calculation, we find the scattering amplitude for helicity-0 massive gravitons to be 
\begin{align} \label{eq:helicity0}
\mathcal{A}_{0000} =&-\frac{1}{576 M_p^2 m^{4N+4}} \left(s^{2 N+3}+t^{2 N+3}+u^{2 N+3}\right) \sum_j \Big( \left(4 c_{1,N,j}+2
   c_{2,N-1,j}+c_{3,N-2,j}\right)^2 \nn \\ 
&+\left(4 d_{1,N-1,j}+2 d_{2,N-2,j}-4 d_{3,N-1,j}-2
   d_{4,N-2,j}-d_{5,N-3,j}\right)^2\Big)+\cdots,
\end{align}
where here and below we display only the leading term for high-energy fixed-angle scattering, i.e. terms with the highest combined power of $s$ and $t$ for $s,t  \gg 1$. 
Considering the leading terms in the amplitude~\eqref{eq:helicity0}, we see that the couplings combine into a sum of squares with the same sign coefficients.  This property follows from the Goldstone equivalence theorem and unitarity of scalar amplitudes. Thus each term in the sum must separately cancel to improve the high-energy growth, since unitarity implies that the couplings are all real.  For each $j$ we thus get the constraints
\begin{align}
c_{3,N-2,j} & = -4 c_{1,N,j}-2 c_{2,N-1,j}, \\
d_{5,N-3,j}& = 4 d_{1,N-1,j}+2 d_{2,N-2,j}-4 d_{3,N-1,j}-2
   d_{4,N-2,j}.
\end{align}

With these constraints imposed, the helicity-$1$ amplitude is now
\begin{align}
\mathcal{A}_{1111} &=-\frac{1}{256 M_p^2 m^{4N}} s^{2 N+1} \sum_j \Big( 4 \left(4 c_{1,N,j}+c_{2,N-1,j}\right)^2+4
   \left(2 d_{1,N-1,j}-4 d_{3,N-1,j}-d_{4,N-2,j}\right)^2 \nn \\
& + \left(2 \tilde{c}_{1,N-1,j}+\tilde{c}_{2,N-2,j}\right)^2+\left(4 \tilde{d}_{1,N-1,j}+2
   \tilde{d}_{2,N-2,j}-2 \tilde{d}_{3,N-2,j}-\tilde{d}_{4,N-3,j}\right)^2 \Big)+\cdots.
\end{align}
This is again a sum of squares, so enforcing that this vanishes gives the additional constraints
\begin{align}
c_{2,N-1,j}& = -4 c_{1,N,j}, \\
d_{4,N-2,j}& = 2 d_{1,N-1,j}-4 d_{3,N-1,j},\\
\tilde{c}_{2,N-2,j} & = -2
   \tilde{c}_{1,N-1,j}, \\
\tilde{d}_{4,N-3,j}&=4 \tilde{d}_{1,N-1,j}+2 \tilde{d}_{2,N-2,j}-2 \tilde{d}_{3,N-2,j}.
\end{align}
With these constraints enforced, the helicity-$2$ amplitude is
\be
\mathcal{A}_{2222} =-\frac{1}{4 M_p^2 m^{4N-4}} s^{2 N-1} \sum_j \left(4
   c_{1,N,j}^2+4 d_{3,N-1,j}^2+\tilde{c}_{1,N-1,j}^2+ \left(2 \tilde{d}_{1,N-1,j}-\tilde{d}_{3,N-2,j}\right)^2\right)+\, \cdots,
\ee
which is again a sum of squares, giving us the further constraints
\be
c_{1,N,j}=d_{3,N-1,j}=\tilde{c}_{1,N-1,j}= 0,\quad \tilde{d}_{3,N-2,j}= 2 \tilde{d}_{1,N-1,j}.
\ee

To constrain the remaining $(2N-1)$-derivative interactions, we need to look at amplitudes with more than one helicity type.\footnote{The leading terms of the two amplitudes we consider next arise at an order $E^2$ lower than the power-counting estimate \eqref{eq:exchangepower}, but they are still more divergent than the corresponding massive gravity terms so have to independently cancel for $N>1$.} The amplitude for helicity-1 and helicity-0 scattering is
\be
\mathcal{A}_{1010} =\frac{  su \left(s^{2 N-1}+u^{2 N-1}\right)}{192 M_p^2 m^{4N+2}} \sum_j m_{A_j}^2 \left(\left(2
   d_{1,N-1,j}+d_{2,N-2,j}\right)^2+4 \tilde{d}_{1,N-1,j}^2\right)+\cdots.
\ee
Setting this to zero gives the constraints
\be
d_{2,N-2,j}= -2 d_{1,N-1,j}, \quad \tilde{d}_{1,N-1,j}= 0.
\ee
Lastly, we look at the amplitude for helicity-2 and helicity-1 scattering, 
\be
\mathcal{A}_{2121} =\frac{1}{32 M_p^2 m^{4N-2}} s^{2 N-2}u\sum_j m_{A_j}^2 \left(\tilde{d}_{2,N-2,j}^2+4 d_{1,N-1,j}^2\right)+\cdots,
\ee
which gives the constraints
\be
\tilde{d}_{2,N-2,j}= d_{1,N-1,j}=0.
\ee

The above argument shows that all of the highest-derivative interactions have to vanish, otherwise the high-energy behavior is at least as bad as in massive gravity. Note that it was important that the leading parts of the amplitudes we considered were not contaminated by contributions from lower-derivative terms.  We can thus repeat this argument for the next highest-derivative interactions, and so on, until only interactions with two or fewer derivatives remain. These remaining interactions contribute at the same order as the pure massive graviton terms, so we next need to check whether these can cancel against each other. 

The helicity-0 amplitude is now
\begin{align}
\mathcal{A}_{0000} =&-\frac{5stu\left(s^2+t^2+u^2\right)}{864 M_p^2 m^8}  \bigg(2 \left(6 c_1+4 c_2-1\right)^2 
\nn\\ & +\sum_j \left(3 \left(2 c_{1,1,j}+c_{2,0,j}\right)^2 +12\left(d_{1,0,j} -d_{3,0,j}\right)^2 \right) \bigg)+\cdots,
\end{align}
which grows like $\sim E^{10}$. We see that the contributions from scalar and vector exchange cannot cancel the pure massive gravity contribution, 
so setting this to zero gives the constraints
\begin{align}
c_2& =\frac{1}{4}-\frac{3}{2}c_1, \\
c_{2,0,j}&= -2 c_{1,1,j}, \\
d_{3,0,j}& = d_{1,0,j}.
\end{align}
Imposing these constraints, the new leading part of the helicity-0 amplitude is
\be
\mathcal{A}_{0000} =-\frac{1}{144 M_p^2 m^6}\left(s^2+t^2+u^2\right)^2 (16 d_1+32 d_3-3)+\cdots.
\ee
Setting this to zero further constrains the coefficients in the graviton potential,
\be
d_3= \frac{3}{32}-\frac{d_1}{2}.
\ee
The constraints we have found on $c_2$ and $d_3$ are the conditions defining the on-shell de Rham-Gabadadze-Tolley (dRGT) potential up to this order~\cite{deRham:2010ik}.
Now we look at the helicity-1 amplitude, whose new leading part is
\be
\mathcal{A}_{1111} = -\frac{1}{64 M_p^2 m^4} s^3 \left(24 (c_1-1)^2+\sum_j \left(4 c_{1,1,j}^2+4
   d_{1,0,j}^2+\tilde{c}_{1,0,j}^2+4 \tilde{d}_{1,0,j}^2\right)\right)+\cdots.
\ee
Requiring that this vanishes, we get the further constraints 
\be
c_{1,1,j}=
   d_{1,0,j}=\tilde{c}_{1,0,j}= \tilde{d}_{1,0,j}=0,\quad c_1=1.
\ee
The remaining helicity-0 amplitude is then 
\be
\mathcal{A}_{0000} =\frac{1}{72 M_p^2 m^4} s t u \left(128 d_1-115-6 \sum_j c_{1,0,j}^2\right)+\cdots,
\ee
and cancelling this gives
\be
d_1 = \frac{1}{128} \left(115+6\sum_j  c_{1,0,j}^2 \right).
\ee

With the conditions determined so far, many but not all of the $\sim E^6$ terms of the amplitudes vanish. One of the surviving amplitudes is
\be
\mathcal{A}_{2000} = \frac{1}{32 \sqrt{6}M_p^2 m^4} s t u \left(1+2 \sum_j c_{1,0,j}^2\right)+\cdots.
\ee
There is no way to set this to zero with real couplings, so we conclude that it is impossible to improve the high-energy behavior for Lagrangians of the form \eqref{eq:Ltot}. This implies that there is no tree-level UV extension of massive gravity with only spin-0 and spin-1 particles.

\section{Model independent no-go result}
\label{sec:part2}

In the previous section we assumed a particular form for the massive spin-2 part of the Lagrangian, namely that it was the Einstein-Hilbert term plus a general potential. Now we relax this assumption and prove that there is no way to improve the high-energy behavior of the tree-level four-point amplitude for \textit{any} theory with a massive spin-2 particle coupled to scalars and vectors.  In addition, we allow for the presence of a single massless spin-2 particle, which covers the case of bigravity models. 

Our approach here is somewhat different than in the previous section.  We bypass the Lagrangian and directly write down the most general four-point amplitude with a given high-energy behavior that is consistent with Lorentz invariance, locality, unitarity, crossing symmetry, and a bounded number of derivatives. We follow the procedure of Refs.~\cite{Bonifacio:2018vzv,Bonifacio:2018aon}, which we review in Appendix~\ref{app:details}. In particular, we construct the amplitudes using general on-shell cubic and quartic vertices.  This encompasses the Lagrangian approach of the previous section as a special case, since any cubic interactions in the Lagrangian that vanish on-shell are equivalent to higher-point interactions under a field redefinition.

The result we derive is stronger than the one from the previous section, but it is also less transparent since we cannot include the lengthy output from the intermediate steps. In Appendix \ref{app:spin1} we consider the simpler example of a single spin-1 particle coupled to scalars and see that improving the high-energy behavior leads to the Abelian Higgs model, as expected.

\subsection{On-shell vertices} \label{sec:dof}

We again consider the coupling of a massive spin-2 particle $h_{\mu \nu}$ to arbitrary numbers of spin-0 particles $\phi_j$ with masses $m_{\phi_j}\geq 0$ and arbitrary numbers of massive spin-1 particles $A_{j}^{\mu}$ with masses $m_{A_j}>0$.  There are no on-shell $h^2 A$ cubic interactions between a massless spin-1 particle and a single real massive spin-2 particle, thus massless spin-1 particles cannot contribute.  In addition, fermions of any spin cannot be exchanged by external bosons due to angular momentum conservation.  Thus the particles we consider are all of the possible degrees of freedom with spins less than two that can contribute to the four-point amplitude with external massive spin-2 particles.  In addition, we now also include couplings to a single massless spin-2 particle, $\gamma_{\mu \nu}$.   

We now list all the relevant on-shell cubic and quartic vertices with these degrees of freedom. Details of how to classify these vertices are given in Appendix~\ref{app:details}.

\subsubsection{Cubic vertices}
Let us start with the cubic vertices. The most general cubic self-interactions of a massive spin-2 particle are described by the following vertex:
\begin{align}
\mathcal{V}_{h^3} &=i a_1 (\epsilon_1 \ccdot \epsilon_2) (\epsilon_1  \ccdot \epsilon_3) (\epsilon_2 \ccdot \epsilon_3) \nn \\ 
&+i a_2 \left( (\epsilon_2 \ccdot \epsilon_3)^2 (\epsilon_1 \ccdot p_2)^2+ (\epsilon_1 \ccdot \epsilon_3)^2 (\epsilon_2 \ccdot p_3)^2+ (\epsilon_1 \ccdot \epsilon_2)^2 (\epsilon_3 \ccdot p_1)^2\right)  \nn \\ & +i a_3 \big(  (\epsilon_1 \ccdot \epsilon_3) (\epsilon_2 \ccdot \epsilon_3) (\epsilon_1 \ccdot p_2) (\epsilon_2 \ccdot p_3)+ (\epsilon_1 \ccdot \epsilon_2) (\epsilon_2 \ccdot \epsilon_3) (\epsilon_1 \ccdot p_2)  (\epsilon_3 \ccdot p_1)+ (\epsilon_1 \ccdot \epsilon_2) (\epsilon_1 \ccdot \epsilon_3) (\epsilon_2 \ccdot p_3)  (\epsilon_3 \ccdot p_1) \big) \nn \\ 
&+i a_4 (\epsilon_1 \ccdot p_2)(\epsilon_2 \ccdot p_3) (\epsilon_3 \ccdot p_1) \big( (\epsilon_1 \ccdot \epsilon_2)(\epsilon_3 \ccdot p_1) +(\epsilon_2 \ccdot \epsilon_3) (\epsilon_1 \ccdot p_2) +(\epsilon_1  \ccdot \epsilon_3) (\epsilon_2 \ccdot p_3) \big) \nn \\ 
&+i a_5 (\epsilon_1 \ccdot p_2)^2 (\epsilon_2 \ccdot p_3)^2 (\epsilon_3 \ccdot p_1)^2 \nn \\
&+ i \tilde{a}_1 \big( (\epsilon_1  \ccdot \epsilon_3) (\epsilon_2  \ccdot \epsilon_3) \varepsilon (p_{1} p_2 \epsilon_{1} \epsilon_2 )- (\epsilon_1 \ccdot \epsilon_2) (\epsilon_2  \ccdot \epsilon_3) \varepsilon (p_{1} p_2 \epsilon_{1} \epsilon_3)+ (\epsilon_1 \ccdot \epsilon_2) (\epsilon_1  \ccdot \epsilon_3) \varepsilon (p_{1} p_2 \epsilon_{2} \epsilon_3) \big) \nn \\
& +i\tilde{a}_2  (\epsilon_1 \ccdot p_2) (\epsilon_2 \ccdot p_3)  (\epsilon_3 \ccdot p_1) \big( (\epsilon_3 \ccdot p_1) \varepsilon ( p_1 p_2 \epsilon_1 \epsilon_2 ) - (\epsilon_2 \ccdot p_3) \varepsilon ( p_1 p_2 \epsilon_1 \epsilon_3)+ (\epsilon_1 \ccdot p_2) \varepsilon ( p_1 p_2 \epsilon_2 \epsilon_3)\big),
\end{align}
where $a_i$ and $\tilde{a}_i$ are (in general dimensionful) coupling constants and $\varepsilon( \cdot)$ denotes the contraction of the antisymmetric tensor with the enclosed vectors.  Due to dimensionally-dependent identities, we can set
\be a_4=0\,\ee
without loss of generality, which from the Lagrangian point of view corresponds to the vanishing of the Gauss-Bonnet term in four dimensions.

The general $h^2 \gamma$ interactions between a massive spin-2 particle and a massless spin-2 particle are described by the following vertex:
\begin{align}
\mathcal{V}_{h^2 \gamma} & = i b_1 (\epsilon_1 \ccdot \epsilon_2)^2 (\epsilon_3 \ccdot p_1)^2 \nn \\ 
&+i b_2 (\epsilon_1 \ccdot \epsilon_2)(\epsilon_3 \ccdot p_1) \big((\epsilon_2 \ccdot \epsilon_3)(\epsilon_1 \ccdot p_2)+(\epsilon_1  \ccdot \epsilon_3)(\epsilon_2 \ccdot p_3) \big) \nn \\
&+i b_3 \big((\epsilon_2 \ccdot \epsilon_3) (\epsilon_1 \ccdot p_2)+(\epsilon_1 \ccdot \epsilon_3)(\epsilon_2 \ccdot p_3) \big)^2 \nn \\ 
&+i b_4 (\epsilon_1 \ccdot p_2) (\epsilon_2 \ccdot p_3) (\epsilon_3 \ccdot p_1) \big((\epsilon_2 \ccdot \epsilon_3) (\epsilon_1 \ccdot p_2)+(\epsilon_1  \ccdot \epsilon_3)(\epsilon_2 \ccdot p_3)\big)\nn \\
& + i b_{5} (\epsilon_1 \ccdot \epsilon_2) (\epsilon_1 \ccdot p_2)(\epsilon_2 \ccdot p_3) (\epsilon_3 \ccdot p_1)^2\nn \\ 
&+i b_{6} (\epsilon_1 \ccdot p_2)^2 (\epsilon_2 \ccdot p_3)^2 (\epsilon_3 \ccdot p_1)^2 \nn \\
& + i \tilde{b}_1 \left( (\epsilon_2  \ccdot \epsilon_3) (\epsilon_1 \ccdot p_2)  + 
  (\epsilon_1  \ccdot \epsilon_3) (\epsilon_2 \ccdot p_3)  \right) \varepsilon (p_3 \epsilon_1 \epsilon_2 \epsilon_3) \nn \\ 
&+ i \tilde{b}_2 (\epsilon_1  \ccdot \epsilon_2) (\epsilon_3 \ccdot p_1) \varepsilon (p_3 \epsilon_1 \epsilon_2 \epsilon_3)  \nn \\
&+ i \tilde{b}_3 (\epsilon_1 \ccdot p_2) (\epsilon_2 \ccdot p_3) (\epsilon_3 \ccdot p_1) \varepsilon (p_3 \epsilon_1 \epsilon_2 \epsilon_3) \nn \\ 
& +i \tilde{b}_4 (\epsilon_1 \ccdot p_2) (\epsilon_2 \ccdot p_3) (\epsilon_3 \ccdot p_1)^2 \varepsilon (p_1 p_2 \epsilon_1 \epsilon_2) ,
\end{align}
where the $b_i$, $\tilde b_i$ are coupling constants and particle 3 is massless.   Using dimensionally-dependent identities, we can set 
\be b_4=0\ee
without loss of generality.  If the massless spin-2 particle self interacts via the cubic Einstein-Hilbert interaction, then gauge invariance implies that\footnote{This follows from gauge invariance of the amplitude for graviton Compton scattering off a massive spin-2 particle, as reviewed in the appendix of Ref.~\cite{Bonifacio:2018aon}, or from consistent factorization of the amplitude in massive spinor-helicity variables~\cite{Chung:2018kqs}. The constraint \eqref{eq:mincoupling} can be violated in theories with linear gauge symmetry~\cite{Bonifacio:2019pfg}, but here we assume that the spin-2 gauge symmetry is nonlinear.}
\be \label{eq:mincoupling}
2 b_1=b_2 = \frac{4}{M_p}.
\ee

The general $h^2 \phi_j$ and $h^2 A_j$ cubic interactions between the massive spin-2 particle and the particles with spin 0 and spin 1 are described by the following vertices:
\begin{align}
\mathcal{V}_{h^2 \phi_j} & = i c_{1,j}  (\epsilon_1 \ccdot \epsilon_2)^2\nn\\
&+ic_{2,j} (\epsilon_1 \ccdot \epsilon_2) (\epsilon_1 \ccdot p_2)(\epsilon_2 \ccdot p_3) \nn\\
&+i c_{3,j}(\epsilon_1 \ccdot p_2)^2 (\epsilon_2 \ccdot p_3)^2 \nn\\
&-i \tilde{c}_{1,j}  (\epsilon_1 \ccdot \epsilon_2)  \varepsilon ( p_1 p_2 \epsilon_1 \epsilon_2 ) \nn\\
&- i \tilde{c}_{2,j} (\epsilon_1 \ccdot p_2) (\epsilon_2 \ccdot p_3)  \varepsilon ( p_1 p_2 \epsilon_1 \epsilon_2 ), \label{eq:scalarcubicint} \\
\mathcal{V}_{h^2 A_j}  &= m_{A_j} \bigg[ d_{1,j}  (\epsilon_1 \ccdot \epsilon_2) \big( (\epsilon_2 \ccdot \epsilon_3) (\epsilon_1 \ccdot p_2)-(\epsilon_1  \ccdot \epsilon_3)(\epsilon_2 \ccdot p_3)\big) \nn\\
&+d_{2,j} (\epsilon_1 \ccdot p_2) (\epsilon_2 \ccdot p_3) \big( (\epsilon_2 \ccdot \epsilon_3) (\epsilon_1 \ccdot p_2)-(\epsilon_1  \ccdot \epsilon_3)(\epsilon_2 \ccdot p_3)\big) \nn \\
& +\tilde{d}_{1,j}  (\epsilon_1 \ccdot \epsilon_2) \big( \varepsilon ( p_1 \epsilon_1 \epsilon_2 \epsilon_3 ) -\varepsilon ( p_2 \epsilon_1 \epsilon_2 \epsilon_3 )\big)\nn\\
& -\tilde{d}_{2,j}\varepsilon ( p_1 p_2 \epsilon_1 \epsilon_2) \big( (\epsilon_2 \ccdot \epsilon_3) (\epsilon_1 \ccdot p_2)-(\epsilon_1  \ccdot \epsilon_3)(\epsilon_2 \ccdot p_3)\big)\bigg]  ,
\end{align}
where $c_{i,j}$, $\tilde{c}_{i,j}$, $d_{i,j}$, and $\tilde{d}_{i,j}$ are coupling constants that are real in a unitary theory and particle 3 is the low-spin particle. These correspond to the on-shell vertices produced by the interactions with $l=0$ in Eq.~\eqref{eq:hhphi} and the first two lines of Eq.~\eqref{eq:hhA} (up to factors of $m$ and $M_p$). The fields can be assigned definite parity if some of the couplings vanish, e.g., a pseudoscalar would have zero $c_{i,j}$ and nonzero $\tilde{c}_{i,j}$. However, we consider the general case where parity is not necessarily conserved. 

\subsubsection{Quartic vertices}

We also need the quartic contact term for identical spin-2 particles. This
can be written as
\be \label{eq:spin2quartic}
\mathcal{V}_{h^4} = i \sum_{I=1}^{201} f_I(s,t) \mathbb{T}_I(\epsilon, p),
\ee
where $f_I(s,t)$ are polynomials in the Mandelstam invariants that we assume have bounded degree. The tensor structures $\mathbb{T}_I(\epsilon, p)$ encode the different ways of contracting polarization tensors and are invariant under the group of permutations that preserve the Mandelstam invariants, $\Pi^{\rm kin}$, which is defined in Eq.~\eqref{eq:kinperms}. 
For example, the six zero-derivative tensor structures are
\begin{subequations}
\begin{align}
\mathbb{T}_1(\epsilon, p) & =(\epsilon_1 \ccdot \epsilon_2)^2 (\epsilon_3 \ccdot \epsilon_4)^2, \\
\mathbb{T}_2(\epsilon, p) & = (\epsilon_1 \ccdot \epsilon_3)^2 (\epsilon_2 \ccdot \epsilon_4)^2, \\
\mathbb{T}_3(\epsilon, p) & =  (\epsilon_1 \ccdot \epsilon_4)^2 (\epsilon_2 \ccdot \epsilon_3)^2, \\
\mathbb{T}_4(\epsilon, p) & =  (\epsilon_1 \ccdot \epsilon_2) (\epsilon_1 \ccdot \epsilon_3)  (\epsilon_2 \ccdot \epsilon_4) (\epsilon_3 \ccdot \epsilon_4) , \\
\mathbb{T}_5(\epsilon, p) & =  (\epsilon_1 \ccdot \epsilon_2) (\epsilon_1 \ccdot \epsilon_4)  (\epsilon_2 \ccdot \epsilon_3) (\epsilon_3 \ccdot \epsilon_4) , \\
\mathbb{T}_6(\epsilon, p) & =  (\epsilon_1 \ccdot \epsilon_3) (\epsilon_1 \ccdot \epsilon_4)  (\epsilon_2 \ccdot \epsilon_3) (\epsilon_2 \ccdot \epsilon_4).
\end{align}
\end{subequations}
Only 97 of the 201 tensor structures we use are independent in four dimensions because of dimensionally-dependent identities, but finding a basis is difficult and not needed for our calculation. We also consider only the parity-even quartic vertex, corresponding to four-point amplitudes where the sum of transversities of the external particles is even, since this is sufficient to prove our result.

\subsection{Results}

With the complete list of on-shell vertices in hand, we can follow the procedure outlined in Appendix~\ref{app:details} to determine if there exists a four-point amplitude with improved high-energy behavior compared to massive gravity.\footnote{An alternative procedure is to construct the fully symmetric contact terms up to some given number of derivatives. We have checked that this gives identical results when including terms with up to 14 derivatives.} The output of this procedure is a system of polynomial equations in the cubic coupling constants and mass ratios.
These equations are sum rules that must be satisfied for the high-energy amplitudes to grow more slowly than $\sim E^6$, similar to what we found in the previous section.

We now show that these equations have no real solutions.  First we look at the constraints that depend only on the coupling constants $a_i$, $\tilde{a}_i$, $b_i$, and $\tilde{b}_i$, which define the self interactions and gravitational interactions of the massive spin-2 particle. Two of these constraints are given by
\be
\tilde{b}_3^2+(b_5-b_6)^2 =0, \quad \tilde{b}_4^2+b_6^2=0,
\ee 
so we conclude that $\tilde{b}_3 = \tilde{b}_4 =b_5=b_6=0$. The remaining constraints of this type are 
\begin{align}
 \left(a_1+2 a_2\right) a_5=\left(2 a_1 -a_5 \right)a_5 =6 \tilde{a}_1 \tilde{a}_2+a_1 a_5 =9 \tilde{a}_2^2+2 a_1 a_5 &=0 ,\\
4 a_2 a_3+\frac{3 a_1 a_5}{8}-\tilde{a}_1^2-4 a_2^2-a_3^2=a_3 a_5 & =0.
\end{align}
The only real solution to these equations is
\be
a_3 = 2a_2, \quad a_5=\tilde{a}_1=\tilde{a}_2=0,
\ee
which correspond to the cubic couplings in dRGT massive gravity.
Substituting this solution into the remaining equations, four of them reduce to
\be
\sum_j c_{3,j}^2 = \sum_j \tilde{c}_{2,j}^2  = \sum_j m_{A_j}^2 d_{2,j}^2= \sum_j m_{A_j}^2 \tilde{d}_{2,j}^2=0,
\ee
so we conclude that
\be
c_{3,j}=\tilde{c}_{2,j}=d_{2,j}=\tilde{d}_{2,j}=0.
\ee
Next we substitute these solutions into the remaining equations and look for linear combinations that depend only on the couplings $b_1, b_2$ and $b_3$. This gives the constraints
\begin{align} \label{eq:bconstraints}
\left(2b_1-b_2\right)b_3 -\tilde{b}_1 \tilde{b}_2= \left(b_1-b_2+b_3\right)b_1=\left(b_1-b_3\right)b_3-\tilde{b}_1^2=b_1^2+3b_1b_3-b_2^2 -\tilde{b}_2^2=0.
\end{align}
The $S$-matrix equivalence principle further implies that $b_2=2b_1$, as in Eq. \eqref{eq:mincoupling}. Enforcing this condition and finding the real solutions to the constraints \eqref{eq:bconstraints} gives
\be
b_2=2b_1 =2 b_3, \quad \tilde{b}_1=\tilde{b}_2=0. 
\ee
By taking appropriate linear combinations of the remaining equations, we get the following additional constraints:
\begin{align}
2a_1^2+3\sum_j \left( c_{2,j}^2 + 4d_{1,j}^2 \right)= 4b_1^2+a_2^2-a_1^2+\sum_j  \left( \tilde{c}_{1,j}^2-2m_{A_j}^2 d_{1,j}^2 + (4+6m_{A_j}^2)\tilde{d}_{1,j}^2 \right) =0.
\end{align}
These imply that 
\be
a_1 =a_2=b_1= c_{2,j} =d_{1,j}=\tilde{c}_{1,j}=\tilde{d}_{1,j}=0.
\ee
Finally, substituting the solutions obtained so far into the remaining equations gives
\be
\sum_j c_{1,j}^2=0,
\ee
so we must also have $c_{1,j}=0$. This shows that all of the cubic couplings must vanish and the only surviving amplitude is the trivial one.

The above argument shows that there are no nontrivial unitary amplitudes that grow more slowly than $\sim E^6$ for high-energy fixed-angle scattering with the degrees of freedom we considered. This rules out the existence of a unitary and Lorentz-invariant tree-level UV extension of any theory with a massive spin-2 particle coupled to gravity and particles with spin less than 2 in four dimensions. Any such UV extension must contain additional massive particles with spins 2 or higher. 

\section{Discussion}

We have shown that it is impossible to improve the high-energy behavior of massive spin-2 tree amplitudes by coupling to particles with spins less than 2, even in the presence of ordinary massless gravity. This implies that any tree-level UV extension of massive gravity or bigravity must include additional massive particles with spins 2 or higher.  This has consequences for many proposed models, e.g. the proposed UV completion of massive gravity with an additional scalar field  of Ref.~\cite{Torabian:2017bqu}.  Indeed, various Lorentz-invariant extensions of massive gravity that include extra scalars, such as quasi-dilaton \cite{D'Amico:2012zv,DeFelice:2013tsa} and galileon-extended models \cite{Gabadadze:2012tr,Andrews:2013ora}, all have high-energy behavior that is the same as in pure massive gravity.

Our result also has consequences for the supersymmetric (SUSY) case.  It might be thought that SUSY could help with the UV behavior of massive gravity.\footnote{Supersymmetry does help in the analogous case of a massive spin-$3/2$ particle. The highest strong coupling scale for a single massive spin-$3/2$ particle coupled to gravity is $\Lambda_2=\left(m M_p \right)^{1/2}$~\cite{Rahman:2011ik}. By including a light scalar and pseudoscalar, this can be raised to $M_p$, as realized by broken $\mathcal{N}=1$ supergravity with a chiral supermultiplet~\cite{Casalbuoni:1988sx}.}  The ${\cal N}=1$ massive spin-2 SUSY multiplet contains a massive spin-2 particle, a massive vector and two massive spin-$3/2$ fermions \cite{Buchbinder:2002gh,Zinoviev:2002xn,Ondo:2016cdv,Zinoviev:2018juc,Garcia-Saenz:2018wnw}.  Since the fermions can never appear in the internal line of the tree-level graviton four-point amplitude, we can restrict attention to only the additional massive vector, which then falls under the assumptions of our result.  

Examples of theories containing massive spin-2 particles that do have improved high-energy behavior come from Kaluza-Klein dimensional reduction. For example, dimensionally reducing 5D General Relativity on a single compact extra dimension gives a lower-dimensional theory containing a tower of complex massive spin-2 particles coupled to gravity, a massless spin-1 graviphoton, and a scalar radion. By 5D momentum conservation, the four-point amplitude of one of these massive spin-2 particles receives contributions from the exchange of the massless fields and a massive spin-2 particle with twice the mass. With these additional fields, cancellations of the worst high-energy parts of the amplitude occur, resulting in a raised strong coupling scale of $\Lambda_{3/2}=\left(m M_p^2 \right)^{1/3}$~\cite{Schwartz:2003vj}.\footnote{In contrast, dimensionally reducing higher-dimensional massive gravity cannot give a theory with a strong coupling scale above $\Lambda_3$~\cite{Bonifacio:2016blz}.}
Another example is string theory, which achieves soft high-energy amplitudes with the exchange of an infinite number of massive higher-spin particles.\footnote{For example,
in open bosonic string theory without Chan-Paton factors, the massive spin-2 particle on the leading Regge trajectory has cubic interactions of the form $2$-$2$-$s$ with all of the even spin-$s$ states on the leading trajectory with masses $m_{s}^2=(s-1)/\alpha'$ \cite{Sagnotti:2010at},
\begin{align}
\mathcal{V}_{2,2,s} &=  i g 2^{ \frac{s}{2}-1}
   (\epsilon_3 \ccdot p_1)^{ 
  s-4}\Big[ (s-3)_4 (\epsilon_1 \ccdot \epsilon_3)^2 (\epsilon_2 \ccdot \epsilon_3)^2 +  4 ( s-2)_3 (\epsilon_1 \ccdot \epsilon_3) (\epsilon_2 \ccdot \epsilon_3) \left( (\epsilon_2 \ccdot \epsilon_3) (\epsilon_1 \ccdot p_2) + (\epsilon_1 \ccdot \epsilon_3) (\epsilon_2 \ccdot p_3)\right) (\epsilon_3 \ccdot p_1) \nn\\
&+ 4 ( s-1) s (2 (\epsilon_1 \ccdot \epsilon_2) (\epsilon_1 \ccdot \epsilon_3) (\epsilon_2 \ccdot \epsilon_3) +  (\epsilon_2 \ccdot \epsilon_3)^2 (\epsilon_1 \ccdot p_2)^2 + 
      4 (\epsilon_1 \ccdot \epsilon_3) (\epsilon_2 \ccdot \epsilon_3) (\epsilon_1 \ccdot p_2) (\epsilon_2 \ccdot p_3)+(\epsilon_1 \ccdot \epsilon_3)^2 (\epsilon_2 \ccdot p_3)^2) (\epsilon_3 \ccdot p_1)^2  \nn \\
&  + 
   16 s \left( (\epsilon_2 \ccdot \epsilon_3) (\epsilon_1 \ccdot p_2)+ (\epsilon_1 \ccdot \epsilon_3) (\epsilon_2 \ccdot p_3)\right) ((\epsilon_1 \ccdot \epsilon_2) + 
      (\epsilon_1 \ccdot p_2)(\epsilon_2 \ccdot p_3)) (\epsilon_3 \ccdot p_1)^3  \nn \\
&+ 
   8 \left((\epsilon_1 \ccdot \epsilon_2)^2 + 4 (\epsilon_1 \ccdot \epsilon_2)(\epsilon_1 \ccdot p_2) (\epsilon_2 \ccdot p_3)+ 
      2 (\epsilon_1 \ccdot p_2)^2 (\epsilon_2 \ccdot p_3)^2\right) (\epsilon_3 \ccdot p_1)^4
\Big],
\end{align}
where $g$ is the string coupling constant and we have set $m_{2}^2=1/\alpha' =1$.
} 
Both of these examples contain infinite towers of massive particles with masses that are not parametrically separated.  An obvious further question is whether there can be a weakly-coupled UV completion of massive gravity with a parametrically large gap between the mass of the graviton and the scale of new physics, or whether any UV extension exists using only a finite number of higher-spin particles.

Finally, we assumed Poincar\'e invariance throughout, but another approach to the low strong coupling scale is to consider backgrounds that break Poincar\'e invariance.  For example, in ghost-free massive gravity there are Poincar\'e violating backgrounds with higher strong coupling scales~\cite{deRham:2016plk}, and in AdS the strong coupling scale is raised and new Higgs-like mechanisms are possible \cite{Karch:2000ct,Karch:2001cw,Porrati:2001gx,Porrati:2001db,Porrati:2003sa,Aharony:2003qf,Duff:2004wh,Kiritsis:2006hy,Aharony:2006hz,deRham:2006pe,Kiritsis:2008at,Apolo:2012gg,Gabadadze:2014rwa,Gabadadze:2015goa,Domokos:2015xka,Gabadadze:2017jom,Bachas:2017rch,Bachas:2018zmb,deRham:2018svs}.
It would be interesting to extend our S-matrix based arguments to AdS by studying the dual CFT correlators.

\vspace{-.2cm}
\paragraph{Acknowledgements:} We would like to thank Brando Bellazzini and Clifford Cheung for helpful conversations.  KH and JB would like the thank the University of Amsterdam for hospitality while this work was completed.
KH acknowledges support from DOE grant DE-SC0019143.   RAR is supported by DOE grant DE-SC0011941 and Simons Foundation Award Number 555117 and by NASA grant NNX16AB27G. 

\appendix

\section{Details and conventions}
\label{app:details}
In this appendix we collect various details and conventions used in our calculations.

\subsection{Kinematics}

Here we specify the kinematics used to calculate four-point scattering amplitudes. We consider center-of-mass scattering of identical particles of mass $m$ in the $xz$-plane with particle 1 incoming along the $+\hat{z}$ direction and particles 3 and 4 outgoing. The momenta can be written as
\be \label{eq:momenta}
p^j_{ \mu} = \left( E, p \sin \theta_j, 0, p \cos \theta_j \right),
\ee
where $j$ labels the external particle, $E^2=p^2+m^2$ and $\theta_1=0$, $\theta_2=\pi$, $\theta_3 = \theta$, $\theta_4 = \theta- \pi$. The Mandelstam variables are defined by
\be
s= -(p_1+p_2)^2, \quad t= -(p_1-p_3)^2, \quad u = -(p_1-p_4)^2. 
\ee
These are related to the center-of-mass energy $E$ and the scattering angle $\theta$ by
\be
s = 4 E^2, \quad \cos \theta = 1 - \frac{2t}{4m^2-s}.
\ee

A massive spin-1 particle has three independent polarization vectors. The standard helicity polarizations used in Section~\ref{sec:part1} are defined by
\begin{align} \label{eq:helicitypol}
\epsilon^{(\pm 1)}_{\mu}(p^j) &= \frac{1}{\sqrt{2}}(0, \mp \cos \theta_j,-i,\pm \sin \theta_j),\\
\epsilon^{(0)}_{\mu}(p^j) &= {1\over m}(p,E \sin \theta_j, 0,E \cos \theta_j),
\end{align}
where $j$ labels the external particle.  These are transverse, orthonormal, and complete. They describe states that have definite values of spin projected in their direction of motion.

To simplify the implementation of crossing symmetry in Section~\ref{sec:part2}, we instead use a basis of polarizations that semi-diagonalize the crossing matrix, the so-called transversity basis~\cite{Kotanski:1965zz, Kotanski:1970,deRham:2017zjm}. For particle $j$, these are given by
\begin{subequations} \label{eq:vecpolz}
\begin{align}
\epsilon^{(\pm 1)}_{ \mu}(p^j) & = \frac{i}{\sqrt{2} m} \left( p,E \sin \theta_j \pm i m\cos \theta_j,0,E \cos \theta_j \mp i m \sin \theta_j \right), \\
\epsilon^{(0)}_{ \mu}(p^j) & = \left( 0,0,1,0 \right).
\end{align}
\end{subequations}
These are transverse, orthonormal, and complete, and describe states with definite spin projection in the direction transverse to the scattering plane. 

A massive spin-2 particle has five polarization tensors. A basis for these can be written in terms of the vector polarizations as
\begin{subequations} \label{eq:tensorpolz}
\begin{align}
\epsilon^{(\pm 2)}_{ \mu \nu} & = \epsilon^{(\pm 1)}_{ \mu} \epsilon^{( \pm 1) }_{\nu}, \\
\epsilon^{(\pm 1)}_{ \mu \nu} & = \frac{1}{\sqrt{2}} \left( \epsilon^{(\pm 1)}_{ \mu} \epsilon^{( 0) }_{\nu}+\epsilon^{(0)}_{ \mu} \epsilon^{( \pm 1) }_{\nu} \right), \\
\epsilon^{(0)}_{ \mu \nu} & = \frac{1}{\sqrt{6}} \left( \epsilon^{(1)}_{ \mu} \epsilon^{( -1) }_{\nu}+ \epsilon^{(-1)}_{ \mu} \epsilon^{( 1) }_{\nu}+2 \epsilon^{(0)}_{ \mu} \epsilon^{( 0) }_{\nu} \right).
\end{align}
\end{subequations}
These are transverse, traceless, orthonormal, and complete. A general polarization can be written as a linear combination of these,
\be
\epsilon_{\mu \nu}^j= \alpha_2^j \epsilon^{(2)}_{\mu \nu}+ \alpha_1^j \epsilon^{(1)}_{\mu \nu}+ \alpha_0^j \epsilon^{(0)}_{\mu \nu}+ \alpha_{-1}^j \epsilon^{(-1)}_{\mu \nu}+ \alpha_{-2}^j \epsilon^{(-2)}_{\mu \nu},
\ee
where
\be
\big|\alpha^j_2 \big|^2+\big|\alpha^j_1\big|^2+\big|\alpha^j_0\big|^2+\big|\alpha^j_{-1}\big|^2+\big|\alpha^j_{-2}\big|^2=1.
\ee

The propagator for a spin-0 particle with mass $m$ is
\be
 \frac{-i}{p^2+m^2 - i\epsilon}.
\ee
Defining the projector
\be
\Pi_{\mu \nu} = \eta_{\mu \nu} + \frac{p_{\mu} p_{\nu}}{m^2},
\ee
 the propagator for a spin-1 particle with mass $m>0 $ is
\be
\frac{-i\Pi_{\mu \nu} }{p^2+m^2 - i\epsilon}\,.
\ee
The propagator for a spin-2 particle with mass $m>0$ is 
\be
-\frac{i}{2}\frac{  \Pi_{\mu_1 \nu_1} \Pi_{\mu_2 \nu_2} + \Pi_{\mu_1  \nu_2} \Pi_{\mu_2 \nu_1} -\frac{2}{3} \Pi_{\mu_1 \mu_2} \Pi_{\nu_1  \nu_2} }{p^2+m^2 - i\epsilon}.
\ee
The massless spin-2 propagator (in de Donder gauge) is
\be
-\frac{i}{2} \frac{ \eta_{\mu_1 \nu_1} \eta_{\mu_2 \nu_2} + \eta_{\mu_1  \nu_2} \eta_{\mu_2 \nu_1} -\eta_{\mu_1 \mu_2} \eta_{\nu_1  \nu_2}}{p^2 - i\epsilon}.
\ee

\subsection{Classifying vertices}
Here we review the classification of on-shell vertices. Consider an $n$-point vertex in $d$ dimensions where particle $i$ has integer spin $s_i$ and mass $m_i$. We write the symmetric polarization tensor $\epsilon^{\mu_1\ldots \mu_{s_i}}_{i}$ formally as a product of vectors $\epsilon^{\mu_1}_i \cdots \epsilon^{\mu_{s_i}}_i$. The vertex can then be written as a polynomial in the Lorentz-invariant contractions $\epsilon_i \cdot \epsilon_j$, $\epsilon_i \cdot p_j$, and $p_i \cdot p_j$, possibly also multiplied by a contraction of the antisymmetric tensor $\varepsilon( \cdot)$ with $\epsilon$'s and $p$'s if $d \leq 2n-1$. These contractions are not independent due to the on-shell conditions 
\be
\epsilon_i \cdot p_i =0, \quad \epsilon_i \cdot \epsilon_i =0, \quad p_i\cdot p_i =-m_i^2,  \quad \sum_{i=1}^n p_i =0.
\ee
Moreover, the amplitude must be linear in each polarization tensor. The tensor structures encoding the possible contractions of polarizations are thus built from the following building blocks \cite{Costa:2011mg}:
\be \label{eq:structures}
\varepsilon(\epsilon_1^{\eta_1}\ldots \epsilon_n^{\eta_n} p_1^{\eta_{n+1}}\ldots p_{n-1}^{\eta_{2n-1}}) \left( \prod_{\substack{i,j=1\\ i <j}}^n (\epsilon_i \cdot \epsilon_j)^{n_{ij}} \right)\left( \prod_{\substack{i,j=1\\ i \neq j, j+1 } }^n(\epsilon_i \cdot p_j)^{m_{ij}}\right),
\ee
where $\eta_i$, $n_{ij}=n_{ji}$, and $m_{ij}$ are nonnegative integers satisfying $0 \leq \eta_i \leq 1$ and
\be
 \sum_{\substack{j=1 \\ j \neq i}}^n n_{i j} +\sum_{\substack{j =1 \\ j \neq i, i+1}}^n m_{i j} +\eta_i  =s_i,
\ee
for $i=1, \ldots, n$.
If $\eta_i=0$ for all $i$, then we drop the $\varepsilon(\cdot)$ factor, otherwise we also require
\be
\sum_{i=1}^{2n-1} \eta_i=d, \quad \sum_{i=1}^{n} \eta_i>0.
\ee
To get a general vertex, each tensor structure is multiplied by a function of the independent contractions of momenta. For a tree-level contact vertex this function is a polynomial. When $d \leq 2n-2$ there can be nonlinear Gram identities, which reduce the number of independent tensor structures.\footnote{Gram identities also reduce the number of independent momenta contractions when $d\leq n-2$.} The number of independent tensor structures can be obtained using the representation theory of stabilizer groups and is equal to the number of independent helicity amplitudes~\cite{Kravchuk:2016qvl, Henning:2017fpj}.

When $n \leq 4$ there can also be fewer independent tensor structures due to permutation symmetries that interchange identical particles without changing the Mandelstam invariants, which are called kinematic permutations~\cite{Kravchuk:2016qvl}. For $n=3$, the symmetry group consists of all permutations of the identical particles, which is the symmetric group $S_{k}$ if $k \leq 3$ particles are identical. For $n=4$, if all external particles are identical then the kinematic permutations are given by a $\mathbb{Z}_2^2$ subgroup of $S_4$~\cite{Kravchuk:2016qvl}, 
\be \label{eq:kinperms}
\Pi^{\rm kin} = \{ \mathcal{I}, (12)(34), (13)(24), (14)(23) \},
\ee
where $\mathcal{I}$ is the identity element. If there are two pairs of identical particles then the symmetry group is $\mathbb{Z}_2$. We always work with tensor structures that are invariant under the kinematic permutations.

Amplitudes with massless external particles must also be gauge invariant.  If particle $j$ is massless then cubic vertices should be invariant under
\be
\epsilon_j \rightarrow \epsilon_j + \xi p_j,
\ee
to first order in $\xi$. For $n>3$, the total amplitude must be gauge invariant.

\subsection{Four-point amplitudes}
\label{app:4pt}

Here we briefly review our procedure for obtaining the general four-point amplitude with a given high-energy scaling, following Refs.~\cite{Bonifacio:2018vzv, Bonifacio:2018aon}.

Denote the four-point tree amplitude for identical external bosons with mass $m$ by $\mathcal{A}_{\tau_1 \tau_2 \tau_3 \tau_4}$, where $\tau_j$ labels the transversity of particle $j$, as given by the polarization basis \eqref{eq:vecpolz}. This can be written as the sum of exchange and contact terms,
\be
\mathcal{A}_{\tau_1 \tau_2 \tau_3 \tau_4} = \mathcal{A}_{\tau_1 \tau_2 \tau_3 \tau_4}^{\rm exchange} +\mathcal{A}_{\tau_1 \tau_2 \tau_3 \tau_4}^{\rm contact},
\ee
where the ambiguity of such a split is unimportant for us.
To calculate the most general $ \mathcal{A}_{\tau_1 \tau_2 \tau_3 \tau_4}$ with a given high-energy scaling $\sim E^n$, we go through the following steps:
\begin{enumerate}
\item Calculate $i \mathcal{A}_{\tau_1 \tau_2 \tau_3 \tau_4}^{\rm exchange}$ using the general cubic vertex for each exchanged particle. 
\item Construct an ansatz for $\mathcal{A}^{\rm contact}$ that factors out the kinematical singularities~\cite{Cohen-Tannoudji:1968lnm,Kotanski1968,Kotanski:1970},
\be
\mathcal{A}^{\rm contact}_{\tau_1 \tau_2 \tau_3 \tau_4}(s,t) = \frac{a^{\rm contact}_{\tau_1 \tau_2 \tau_3 \tau_4}(s, t)+i \sqrt{s t u} \, b^{\rm contact}_{\tau_1 \tau_2 \tau_3 \tau_4}( s,t)}{\left(s-4m^2\right)^{ |\sum_i \tau_i|/2}},
\ee
where $a^{\rm contact}_{\tau_1 \tau_2 \tau_3 \tau_4}(s, t)$ and $b^{\rm contact}_{\tau_1 \tau_2 \tau_3 \tau_4}(s, t)$ are polynomials to be determined. 
\item Constrain the above polynomials by the requirement that they cancel the exchange terms when the total amplitude is expanded at high energies, down to whatever assumed high-energy scaling is taken as input. Replace the products of cubic couplings and masses with new variables so that the equations are linear.
\item Impose crossing symmetry on the contact terms~\cite{Kotanski:1965zz, deRham:2017zjm}:
\begin{align}
\mathcal{A}^{\rm contact}_{\tau_1 \tau_2 \tau_3 \tau_4}(s,t) & = e^{i\left(\pi- \chi_t\right) \sum_j \tau_j }\mathcal{A}^{\rm contact}_{-\tau_1 -\tau_3 -\tau_2 -\tau_4}(t,s),\\
\mathcal{A}^{\rm contact}_{\tau_1 \tau_2 \tau_3 \tau_4}(s,t) & = e^{i \left( \pi- \chi_u\right) \sum_j \tau_j }\mathcal{A}^{\rm contact}_{-\tau_1 -\tau_4 -\tau_3 -\tau_2}(u,t),
\end{align}
where
\begin{align}
e^{-i \chi_t} \equiv \frac{-st -2 i m \sqrt{s t u}}{\sqrt{ s(s-4m^2)t(t-4m^2)}}, \quad e^{-i \chi_u} \equiv \frac{-su +2 im \sqrt{s t u}}{\sqrt{ s(s-4m^2)u(u-4m^2)}}.
\end{align} 
These can be cast as linear equations in the parameters by equating the coefficients of the monomials in $s$, $t$, and $\sqrt{s t u}$ on each side.
\item Impose little group covariance by enforcing that $i\mathcal{A}_{\tau_1 \tau_2 \tau_3 \tau_4}^{\rm contact}$ matches a covariant quartic vertex evaluated at four-dimensional kinematics. 
\item Solve the nonlinear equations relating the products of cubic couplings and masses to the linear variables defined earlier.
The parameters from the contact terms appear linearly and are easily eliminated, so the result is a system of polynomial equations in the cubic coupling constants and mass ratios.
\end{enumerate}

\section{Massive spin-1 example}
\label{app:spin1}

In this appendix we apply the procedure of Section~\ref{sec:part2} to the simple example of massive spin-1 scattering with scalar exchange, verifying that we find the expected Abelian Higgs model. 
We compute the four-point amplitude where all external particles have spin 1. For a single spin-1 particle, the best nontrivial high-energy behavior of this amplitude is $\sim E^4$, so we look for amplitudes that grow more slowly than this. 

First we need to write down all the relevant vertices. Our degrees of freedom are a single massive vector, $A_{\mu}$, with mass $m_A$ and a collection of real scalars, $\phi_j$, with masses $m_{\phi_j}$. 
The general on-shell cubic vertex between $A_{\mu}$ and $\phi_j$ that contributes to the four-point spin-1 amplitude is
\begin{align}
\mathcal{V}_{A^2 \phi_j} &=i m_A^2 g_{1,j} \epsilon_{1} \ccdot \epsilon_{2} +i g_{2,j} (\epsilon_1 \ccdot p_2) (\epsilon_2 \ccdot p_3 ) + i \tilde{g}_{1,j} \varepsilon (p_{1} p_2 \epsilon_{1} \epsilon_2),
\end{align}
where $g_{1,j}$, $g_{2,j}$, and $\tilde{g}_{3,j}$ are real coupling constants.
There is no on-shell cubic self-interaction for a single spin-1 particle.
The general quartic vertex with external spin-1 particles is
\be \label{eq:spin1quartic}
\mathcal{V}_{A^4} = i \sum_{I=1}^{17} f_I(s,t) \mathbb{T}_I(\epsilon, p),
\ee
where $f_i(s,t)$ are polynomials in the Mandelstam variables and $\mathbb{T}_I(\epsilon, p)$ are $\mathbb{Z}_2^2$-invariant tensor structures. A basis for these structures can be found in Appendix A of Ref.~\cite{Bonifacio:2018vzv}.

Applying the procedure outlined in \ref{app:4pt}, we find that it is possible to reduce the high-energy behavior of the amplitude to $\sim E^2$.  However, with the appropriate contact terms added, this gives no constraints on the cubic couplings.   We can further improve the high-energy behavior if the following sum rules are satisfied:
\begin{align}
\sum_j g_{2,j} \left(g_{2,j} (m_{\phi_j}^2-2m_A^2)+2 g_{1,j}\right) &=0,  \label{eq:vectorcons1} \\
\sum_j \left( g_{2,j}^2+\tilde{g}_{1,j}^2 \right) &=0.  \label{eq:vectorcons2}
\end{align}
The only real solution to these equations is
\be
g_{2,j}=\tilde{g}_{1,j}=0.
\ee
This corresponds to the Abelian Higgs theory.
The remaining amplitudes are then bounded at high energies by constants depending on the cubic couplings $g_{1,j}$ and masses $m_{\phi_j}$. Perturbative unitarity implies that these constants cannot be too large, so there are further constraints on the masses of the spin-0 particles, as in the Lee-Quigg-Thacker bound on the Higgs mass~\cite{Lee:1977eg}. 

\bibliographystyle{utphys}
\addcontentsline{toc}{section}{References}
\bibliography{Higgs-PRD2}

\providecommand{\href}[2]{#2}\begingroup\raggedright\begin{thebibliography}{10}

\bibitem{ArkaniHamed:2003uy}
N.~Arkani-Hamed, H.-C. Cheng, M.~A. Luty, and S.~Mukohyama, ``{Ghost
  condensation and a consistent infrared modification of gravity},''
  \href{http://dx.doi.org/10.1088/1126-6708/2004/05/074}{{\em JHEP} {\bfseries
  05} (2004) 074},
\href{http://arxiv.org/abs/hep-th/0312099}{{\ttfamily arXiv:hep-th/0312099
  [hep-th]}}.
%%CITATION = HEP-TH/0312099;%%.

\bibitem{Percacci:1990wy}
R.~Percacci, ``{The Higgs phenomenon in quantum gravity},''
  \href{http://dx.doi.org/10.1016/0550-3213(91)90510-5}{{\em Nucl. Phys.}
  {\bfseries B353} (1991) 271--290},
\href{http://arxiv.org/abs/0712.3545}{{\ttfamily arXiv:0712.3545 [hep-th]}}.
%%CITATION = ARXIV:0712.3545;%%.

\bibitem{Kakushadze:2000zn}
Z.~Kakushadze and P.~Langfelder, ``{Gravitational Higgs mechanism},''
  \href{http://dx.doi.org/10.1142/S0217732300002693}{{\em Mod. Phys. Lett.}
  {\bfseries A15} (2000) 2265--2280},
\href{http://arxiv.org/abs/hep-th/0011245}{{\ttfamily arXiv:hep-th/0011245
  [hep-th]}}.
%%CITATION = HEP-TH/0011245;%%.

\bibitem{Chamseddine:2003ft}
A.~H. Chamseddine, ``{Spontaneous symmetry breaking for massive spin-2
  interacting with gravity},''
  \href{http://dx.doi.org/10.1016/S0370-2693(03)00190-4}{{\em Phys. Lett.}
  {\bfseries B557} (2003) 247--252},
\href{http://arxiv.org/abs/hep-th/0301014}{{\ttfamily arXiv:hep-th/0301014
  [hep-th]}}.
%%CITATION = HEP-TH/0301014;%%.

\bibitem{Bandos:2003tm}
I.~A. Bandos, J.~A. de~Azcarraga, J.~M. Izquierdo, and J.~Lukierski,
  ``{Gravity, p-branes and a space-time counterpart of the Higgs effect},''
  \href{http://dx.doi.org/10.1103/PhysRevD.68.046004}{{\em Phys. Rev.}
  {\bfseries D68} (2003) 046004},
\href{http://arxiv.org/abs/hep-th/0301255}{{\ttfamily arXiv:hep-th/0301255
  [hep-th]}}.
%%CITATION = HEP-TH/0301255;%%.

\bibitem{Kirsch:2005st}
I.~Kirsch, ``{A Higgs mechanism for gravity},''
  \href{http://dx.doi.org/10.1103/PhysRevD.72.024001}{{\em Phys. Rev.}
  {\bfseries D72} (2005) 024001},
\href{http://arxiv.org/abs/hep-th/0503024}{{\ttfamily arXiv:hep-th/0503024
  [hep-th]}}.
%%CITATION = HEP-TH/0503024;%%.

\bibitem{Leclerc:2005qc}
M.~Leclerc, ``{The Higgs sector of gravitational gauge theories},''
  \href{http://dx.doi.org/10.1016/j.aop.2005.08.009}{{\em Annals Phys.}
  {\bfseries 321} (2006) 708--743},
\href{http://arxiv.org/abs/gr-qc/0502005}{{\ttfamily arXiv:gr-qc/0502005
  [gr-qc]}}.
%%CITATION = GR-QC/0502005;%%.

\bibitem{tHooft:2007rwo}
G.~'t~Hooft, ``{Unitarity in the Brout-Englert-Higgs Mechanism for Gravity},''
\href{http://arxiv.org/abs/0708.3184}{{\ttfamily arXiv:0708.3184 [hep-th]}}.
%%CITATION = ARXIV:0708.3184;%%.

\bibitem{Kakushadze:2007hf}
Z.~Kakushadze, ``{Massive Gravity in Minkowski Space via Gravitational Higgs
  Mechanism},'' \href{http://dx.doi.org/10.1103/PhysRevD.77.024001}{{\em Phys.
  Rev.} {\bfseries D77} (2008) 024001},
\href{http://arxiv.org/abs/0710.1061}{{\ttfamily arXiv:0710.1061 [hep-th]}}.
%%CITATION = ARXIV:0710.1061;%%.

\bibitem{Wever:2009laa}
C.~S.~P. Wever, ``{A Higgs Mechanism for Gravity},'' Master's thesis, Utrecht
  U., 2009-01-19.
\newblock
  \url{http://web.science.uu.nl/drstp/SHELL/2009/Theses/thesisWever.pdf}.

\bibitem{Pirinccioglu:2009bc}
N.~Pirinccioglu, ``{Gravitational Higgs Mechanism: The Role of Determinantal
  Invariants},'' \href{http://dx.doi.org/10.1007/s10714-012-1414-8}{{\em Gen.
  Rel. Grav.} {\bfseries 44} (2012) 2563--2570},
\href{http://arxiv.org/abs/0908.3367}{{\ttfamily arXiv:0908.3367 [gr-qc]}}.
%%CITATION = ARXIV:0908.3367;%%.

\bibitem{Chamseddine:2010ub}
A.~H. Chamseddine and V.~Mukhanov, ``{Higgs for Graviton: Simple and Elegant
  Solution},'' \href{http://dx.doi.org/10.1007/JHEP08(2010)011}{{\em JHEP}
  {\bfseries 08} (2010) 011},
\href{http://arxiv.org/abs/1002.3877}{{\ttfamily arXiv:1002.3877 [hep-th]}}.
%%CITATION = ARXIV:1002.3877;%%.

\bibitem{Oda:2010wn}
I.~Oda, ``{Higgs Mechanism for Gravitons},''
  \href{http://dx.doi.org/10.1142/S0217732310033724}{{\em Mod. Phys. Lett.}
  {\bfseries A25} (2010) 2411--2421},
\href{http://arxiv.org/abs/1003.1437}{{\ttfamily arXiv:1003.1437 [hep-th]}}.
%%CITATION = ARXIV:1003.1437;%%.

\bibitem{Oda:2010gn}
I.~Oda, ``{Remarks on Higgs Mechanism for Gravitons},''
  \href{http://dx.doi.org/10.1016/j.physletb.2010.05.048}{{\em Phys. Lett.}
  {\bfseries B690} (2010) 322--327},
\href{http://arxiv.org/abs/1004.3078}{{\ttfamily arXiv:1004.3078 [hep-th]}}.
%%CITATION = ARXIV:1004.3078;%%.

\bibitem{Berezhiani:2010xy}
L.~Berezhiani and M.~Mirbabayi, ``{Unitarity Check in Gravitational Higgs
  Mechanism},'' \href{http://dx.doi.org/10.1103/PhysRevD.83.067701}{{\em Phys.
  Rev.} {\bfseries D83} (2011) 067701},
\href{http://arxiv.org/abs/1010.3288}{{\ttfamily arXiv:1010.3288 [hep-th]}}.
%%CITATION = ARXIV:1010.3288;%%.

\bibitem{Iglesias:2011it}
A.~Iglesias and Z.~Kakushadze, ``{Non-perturbative Unitarity of Gravitational
  Higgs Mechanism},'' \href{http://dx.doi.org/10.1103/PhysRevD.84.084005}{{\em
  Phys. Rev.} {\bfseries D84} (2011) 084005},
\href{http://arxiv.org/abs/1102.4991}{{\ttfamily arXiv:1102.4991 [hep-th]}}.
%%CITATION = ARXIV:1102.4991;%%.

\bibitem{Blas:2014ira}
D.~Blas and S.~Sibiryakov, ``{Completing Lorentz violating massive gravity at
  high energies},'' \href{http://dx.doi.org/10.7868/S0044451015030180,
  10.1134/S1063776115030164}{{\em Zh. Eksp. Teor. Fiz.} {\bfseries 147} (2015)
  578--594}, \href{http://arxiv.org/abs/1410.2408}{{\ttfamily arXiv:1410.2408
  [hep-th]}}.
[J. Exp. Theor. Phys.120,no.3,509(2015)].
%%CITATION = ARXIV:1410.2408;%%.

\bibitem{Caron-Huot:2016icg}
S.~Caron-Huot, Z.~Komargodski, A.~Sever, and A.~Zhiboedov, ``{Strings from
  Massive Higher Spins: The Asymptotic Uniqueness of the Veneziano
  Amplitude},'' \href{http://dx.doi.org/10.1007/JHEP10(2017)026}{{\em JHEP}
  {\bfseries 10} (2017) 026},
\href{http://arxiv.org/abs/1607.04253}{{\ttfamily arXiv:1607.04253 [hep-th]}}.
%%CITATION = ARXIV:1607.04253;%%.

\bibitem{Torabian:2017bqu}
M.~Torabian, ``{dRGT theory of massive gravity from spontaneous symmetry
  breaking},'' \href{http://dx.doi.org/10.1016/j.physletb.2018.02.048}{{\em
  Phys. Lett.} {\bfseries B780} (2018) 81--85},
\href{http://arxiv.org/abs/1707.04403}{{\ttfamily arXiv:1707.04403 [hep-th]}}.
%%CITATION = ARXIV:1707.04403;%%.

\bibitem{Bonifacio:2018vzv}
J.~Bonifacio and K.~Hinterbichler, ``{Bounds on Amplitudes in Effective
  Theories with Massive Spinning Particles},''
  \href{http://dx.doi.org/10.1103/PhysRevD.98.045003}{{\em Phys. Rev.}
  {\bfseries D98} no.~4, (2018) 045003},
\href{http://arxiv.org/abs/1804.08686}{{\ttfamily arXiv:1804.08686 [hep-th]}}.
%%CITATION = ARXIV:1804.08686;%%.

\bibitem{Christensen:2014wra}
N.~D. Christensen and Stefanus, ``{On Tree-Level Unitarity in Theories of
  Massive Spin-2 Bosons},''
\href{http://arxiv.org/abs/1407.0438}{{\ttfamily arXiv:1407.0438 [hep-ph]}}.
%%CITATION = ARXIV:1407.0438;%%.

\bibitem{Arkani-Hamed:2017jhn}
N.~Arkani-Hamed, T.-C. Huang, and Y.-t. Huang, ``{Scattering Amplitudes For All
  Masses and Spins},''
\href{http://arxiv.org/abs/1709.04891}{{\ttfamily arXiv:1709.04891 [hep-th]}}.
%%CITATION = ARXIV:1709.04891;%%.

\bibitem{deRham:2018qqo}
C.~de~Rham, S.~Melville, A.~J. Tolley, and S.-Y. Zhou, ``{Positivity Bounds for
  Massive Spin-1 and Spin-2 Fields},''
\href{http://arxiv.org/abs/1804.10624}{{\ttfamily arXiv:1804.10624 [hep-th]}}.
%%CITATION = ARXIV:1804.10624;%%.

\bibitem{Adams:2006sv}
A.~Adams, N.~Arkani-Hamed, S.~Dubovsky, A.~Nicolis, and R.~Rattazzi,
  ``{Causality, analyticity and an IR obstruction to UV completion},''
  \href{http://dx.doi.org/10.1088/1126-6708/2006/10/014}{{\em JHEP} {\bfseries
  10} (2006) 014},
\href{http://arxiv.org/abs/hep-th/0602178}{{\ttfamily arXiv:hep-th/0602178
  [hep-th]}}.
%%CITATION = HEP-TH/0602178;%%.

\bibitem{Cheung:2016yqr}
C.~Cheung and G.~N. Remmen, ``{Positive Signs in Massive Gravity},''
  \href{http://dx.doi.org/10.1007/JHEP04(2016)002}{{\em JHEP} {\bfseries 04}
  (2016) 002},
\href{http://arxiv.org/abs/1601.04068}{{\ttfamily arXiv:1601.04068 [hep-th]}}.
%%CITATION = ARXIV:1601.04068;%%.

\bibitem{Camanho:2016opx}
X.~O. Camanho, G.~Lucena~Gomez, and R.~Rahman, ``{Causality Constraints on
  Massive Gravity},'' \href{http://dx.doi.org/10.1103/PhysRevD.96.084007}{{\em
  Phys. Rev.} {\bfseries D96} no.~8, (2017) 084007},
\href{http://arxiv.org/abs/1610.02033}{{\ttfamily arXiv:1610.02033 [hep-th]}}.
%%CITATION = ARXIV:1610.02033;%%.

\bibitem{Bellazzini:2016xrt}
B.~Bellazzini, ``{Softness and amplitudes' positivity for spinning
  particles},'' \href{http://dx.doi.org/10.1007/JHEP02(2017)034}{{\em JHEP}
  {\bfseries 02} (2017) 034},
\href{http://arxiv.org/abs/1605.06111}{{\ttfamily arXiv:1605.06111 [hep-th]}}.
%%CITATION = ARXIV:1605.06111;%%.

\bibitem{Bellazzini:2017fep}
B.~Bellazzini, F.~Riva, J.~Serra, and F.~Sgarlata, ``{Beyond Positivity Bounds
  and the Fate of Massive Gravity},''
  \href{http://dx.doi.org/10.1103/PhysRevLett.120.161101}{{\em Phys. Rev.
  Lett.} {\bfseries 120} no.~16, (2018) 161101},
\href{http://arxiv.org/abs/1710.02539}{{\ttfamily arXiv:1710.02539 [hep-th]}}.
%%CITATION = ARXIV:1710.02539;%%.

\bibitem{deRham:2017zjm}
C.~de~Rham, S.~Melville, A.~J. Tolley, and S.-Y. Zhou, ``{UV complete me:
  Positivity Bounds for Particles with Spin},''
  \href{http://dx.doi.org/10.1007/JHEP03(2018)011}{{\em JHEP} {\bfseries 03}
  (2018) 011},
\href{http://arxiv.org/abs/1706.02712}{{\ttfamily arXiv:1706.02712 [hep-th]}}.
%%CITATION = ARXIV:1706.02712;%%.

\bibitem{deRham:2017xox}
C.~de~Rham, S.~Melville, and A.~J. Tolley, ``{Improved Positivity Bounds and
  Massive Gravity},'' \href{http://dx.doi.org/10.1007/JHEP04(2018)083}{{\em
  JHEP} {\bfseries 04} (2018) 083},
\href{http://arxiv.org/abs/1710.09611}{{\ttfamily arXiv:1710.09611 [hep-th]}}.
%%CITATION = ARXIV:1710.09611;%%.

\bibitem{Hinterbichler:2017qyt}
K.~Hinterbichler, A.~Joyce, and R.~A. Rosen, ``{Massive Spin-2 Scattering and
  Asymptotic Superluminality},''
  \href{http://dx.doi.org/10.1007/JHEP03(2018)051}{{\em JHEP} {\bfseries 03}
  (2018) 051},
\href{http://arxiv.org/abs/1708.05716}{{\ttfamily arXiv:1708.05716 [hep-th]}}.
%%CITATION = ARXIV:1708.05716;%%.

\bibitem{Bonifacio:2017nnt}
J.~Bonifacio, K.~Hinterbichler, A.~Joyce, and R.~A. Rosen, ``{Massive and
  Massless Spin-2 Scattering and Asymptotic Superluminality},''
  \href{http://dx.doi.org/10.1007/JHEP06(2018)075}{{\em JHEP} {\bfseries 06}
  (2018) 075},
\href{http://arxiv.org/abs/1712.10020}{{\ttfamily arXiv:1712.10020 [hep-th]}}.
%%CITATION = ARXIV:1712.10020;%%.

\bibitem{Afkhami-Jeddi:2018apj}
N.~Afkhami-Jeddi, S.~Kundu, and A.~Tajdini, ``{A Bound on Massive Higher Spin
  Particles},''
\href{http://arxiv.org/abs/1811.01952}{{\ttfamily arXiv:1811.01952 [hep-th]}}.
%%CITATION = ARXIV:1811.01952;%%.

\bibitem{deRham:2010ik}
C.~de~Rham and G.~Gabadadze, ``{Generalization of the Fierz-Pauli Action},''
  \href{http://dx.doi.org/10.1103/PhysRevD.82.044020}{{\em Phys. Rev.}
  {\bfseries D82} (2010) 044020},
\href{http://arxiv.org/abs/1007.0443}{{\ttfamily arXiv:1007.0443 [hep-th]}}.
%%CITATION = ARXIV:1007.0443;%%.

\bibitem{deRham:2010kj}
C.~de~Rham, G.~Gabadadze, and A.~J. Tolley, ``{Resummation of Massive
  Gravity},'' \href{http://dx.doi.org/10.1103/PhysRevLett.106.231101}{{\em
  Phys. Rev. Lett.} {\bfseries 106} (2011) 231101},
\href{http://arxiv.org/abs/1011.1232}{{\ttfamily arXiv:1011.1232 [hep-th]}}.
%%CITATION = ARXIV:1011.1232;%%.

\bibitem{Hassan:2011hr}
S.~F. Hassan and R.~A. Rosen, ``{Resolving the Ghost Problem in non-Linear
  Massive Gravity},''
  \href{http://dx.doi.org/10.1103/PhysRevLett.108.041101}{{\em Phys. Rev.
  Lett.} {\bfseries 108} (2012) 041101},
\href{http://arxiv.org/abs/1106.3344}{{\ttfamily arXiv:1106.3344 [hep-th]}}.
%%CITATION = ARXIV:1106.3344;%%.

\bibitem{Hinterbichler:2011tt}
K.~Hinterbichler, ``{Theoretical Aspects of Massive Gravity},''
  \href{http://dx.doi.org/10.1103/RevModPhys.84.671}{{\em Rev. Mod. Phys.}
  {\bfseries 84} (2012) 671--710},
\href{http://arxiv.org/abs/1105.3735}{{\ttfamily arXiv:1105.3735 [hep-th]}}.
%%CITATION = ARXIV:1105.3735;%%.

\bibitem{deRham:2014zqa}
C.~de~Rham, ``{Massive Gravity},''
  \href{http://dx.doi.org/10.12942/lrr-2014-7}{{\em Living Rev. Rel.}
  {\bfseries 17} (2014) 7},
\href{http://arxiv.org/abs/1401.4173}{{\ttfamily arXiv:1401.4173 [hep-th]}}.
%%CITATION = ARXIV:1401.4173;%%.

\bibitem{Hassan:2011zd}
S.~F. Hassan and R.~A. Rosen, ``{Bimetric Gravity from Ghost-free Massive
  Gravity},'' \href{http://dx.doi.org/10.1007/JHEP02(2012)126}{{\em JHEP}
  {\bfseries 02} (2012) 126},
\href{http://arxiv.org/abs/1109.3515}{{\ttfamily arXiv:1109.3515 [hep-th]}}.
%%CITATION = ARXIV:1109.3515;%%.

\bibitem{Hinterbichler:2012cn}
K.~Hinterbichler and R.~A. Rosen, ``{Interacting Spin-2 Fields},''
  \href{http://dx.doi.org/10.1007/JHEP07(2012)047}{{\em JHEP} {\bfseries 07}
  (2012) 047},
\href{http://arxiv.org/abs/1203.5783}{{\ttfamily arXiv:1203.5783 [hep-th]}}.
%%CITATION = ARXIV:1203.5783;%%.

\bibitem{ArkaniHamed:2002sp}
N.~Arkani-Hamed, H.~Georgi, and M.~D. Schwartz, ``{Effective field theory for
  massive gravitons and gravity in theory space},''
  \href{http://dx.doi.org/10.1016/S0003-4916(03)00068-X}{{\em Annals Phys.}
  {\bfseries 305} (2003) 96--118},
\href{http://arxiv.org/abs/hep-th/0210184}{{\ttfamily arXiv:hep-th/0210184
  [hep-th]}}.
%%CITATION = HEP-TH/0210184;%%.

\bibitem{Schwartz:2003vj}
M.~D. Schwartz, ``{Constructing gravitational dimensions},''
  \href{http://dx.doi.org/10.1103/PhysRevD.68.024029}{{\em Phys. Rev.}
  {\bfseries D68} (2003) 024029},
\href{http://arxiv.org/abs/hep-th/0303114}{{\ttfamily arXiv:hep-th/0303114
  [hep-th]}}.
%%CITATION = HEP-TH/0303114;%%.

\bibitem{Bonifacio:2018aon}
J.~Bonifacio and K.~Hinterbichler, ``{Universal bound on the strong coupling
  scale of a gravitationally coupled massive spin-2 particle},''
  \href{http://dx.doi.org/10.1103/PhysRevD.98.085006}{{\em Phys. Rev.}
  {\bfseries D98} no.~8, (2018) 085006},
\href{http://arxiv.org/abs/1806.10607}{{\ttfamily arXiv:1806.10607 [hep-th]}}.
%%CITATION = ARXIV:1806.10607;%%.

\bibitem{Chung:2018kqs}
M.-Z. Chung, Y.-T. Huang, J.-W. Kim, and S.~Lee, ``{The simplest massive
  S-matrix: from minimal coupling to Black Holes},''
\href{http://arxiv.org/abs/1812.08752}{{\ttfamily arXiv:1812.08752 [hep-th]}}.
%%CITATION = ARXIV:1812.08752;%%.

\bibitem{Bonifacio:2019pfg}
J.~Bonifacio, K.~Hinterbichler, and L.~A. Johnson, ``{Pseudolinear spin-2
  interactions},'' \href{http://dx.doi.org/10.1103/PhysRevD.99.024037}{{\em
  Phys. Rev.} {\bfseries D99} no.~2, (2019) 024037},
\href{http://arxiv.org/abs/1806.00483}{{\ttfamily arXiv:1806.00483 [hep-th]}}.
%%CITATION = ARXIV:1806.00483;%%.

\bibitem{D'Amico:2012zv}
G.~D'Amico, G.~Gabadadze, L.~Hui, and D.~Pirtskhalava, ``{Quasidilaton: Theory
  and cosmology},'' \href{http://dx.doi.org/10.1103/PhysRevD.87.064037}{{\em
  Phys. Rev.} {\bfseries D87} (2013) 064037},
\href{http://arxiv.org/abs/1206.4253}{{\ttfamily arXiv:1206.4253 [hep-th]}}.
%%CITATION = ARXIV:1206.4253;%%.

\bibitem{DeFelice:2013tsa}
A.~De~Felice and S.~Mukohyama, ``{Towards consistent extension of quasidilaton
  massive gravity},''
  \href{http://dx.doi.org/10.1016/j.physletb.2013.12.041}{{\em Phys. Lett.}
  {\bfseries B728} (2014) 622--625},
\href{http://arxiv.org/abs/1306.5502}{{\ttfamily arXiv:1306.5502 [hep-th]}}.
%%CITATION = ARXIV:1306.5502;%%.

\bibitem{Gabadadze:2012tr}
G.~Gabadadze, K.~Hinterbichler, J.~Khoury, D.~Pirtskhalava, and M.~Trodden,
  ``{A Covariant Master Theory for Novel Galilean Invariant Models and Massive
  Gravity},'' \href{http://dx.doi.org/10.1103/PhysRevD.86.124004}{{\em Phys.
  Rev.} {\bfseries D86} (2012) 124004},
\href{http://arxiv.org/abs/1208.5773}{{\ttfamily arXiv:1208.5773 [hep-th]}}.
%%CITATION = ARXIV:1208.5773;%%.

\bibitem{Andrews:2013ora}
M.~Andrews, G.~Goon, K.~Hinterbichler, J.~Stokes, and M.~Trodden, ``{Massive
  Gravity Coupled to Galileons is Ghost-Free},''
  \href{http://dx.doi.org/10.1103/PhysRevLett.111.061107}{{\em Phys. Rev.
  Lett.} {\bfseries 111} no.~6, (2013) 061107},
\href{http://arxiv.org/abs/1303.1177}{{\ttfamily arXiv:1303.1177 [hep-th]}}.
%%CITATION = ARXIV:1303.1177;%%.

\bibitem{Rahman:2011ik}
R.~Rahman, ``{Helicity-1/2 mode as a probe of interactions of a massive
  Rarita-Schwinger field},''
  \href{http://dx.doi.org/10.1103/PhysRevD.87.065030}{{\em Phys. Rev.}
  {\bfseries D87} no.~6, (2013) 065030},
\href{http://arxiv.org/abs/1111.3366}{{\ttfamily arXiv:1111.3366 [hep-th]}}.
%%CITATION = ARXIV:1111.3366;%%.

\bibitem{Casalbuoni:1988sx}
R.~Casalbuoni, S.~De~Curtis, D.~Dominici, F.~Feruglio, and R.~Gatto, ``{When
  does supergravity become strong?},''
  \href{http://dx.doi.org/10.1016/0370-2693(89)91123-4}{{\em Phys. Lett.}
  {\bfseries B216} (1989) 325}.
[Erratum: Phys. Lett.B229,439(1989)].
%%CITATION = PHLTA,B216,325;%%.

\bibitem{Buchbinder:2002gh}
I.~L. Buchbinder, S.~J. Gates, Jr., W.~D. Linch, III, and J.~Phillips, ``{New
  4-D, N=1 superfield theory: Model of free massive superspin 3/2 multiplet},''
  \href{http://dx.doi.org/10.1016/S0370-2693(02)01772-0}{{\em Phys. Lett.}
  {\bfseries B535} (2002) 280--288},
\href{http://arxiv.org/abs/hep-th/0201096}{{\ttfamily arXiv:hep-th/0201096
  [hep-th]}}.
%%CITATION = HEP-TH/0201096;%%.

\bibitem{Zinoviev:2002xn}
{\relax Yu}.~M. Zinoviev, ``{Massive spin two supermultiplets},''
\href{http://arxiv.org/abs/hep-th/0206209}{{\ttfamily arXiv:hep-th/0206209
  [hep-th]}}.
%%CITATION = HEP-TH/0206209;%%.

\bibitem{Ondo:2016cdv}
N.~A. Ondo and A.~J. Tolley, ``{Deconstructing Supergravity: Massive
  Supermultiplets},'' \href{http://dx.doi.org/10.1007/JHEP11(2018)082}{{\em
  JHEP} {\bfseries 11} (2018) 082},
\href{http://arxiv.org/abs/1612.08752}{{\ttfamily arXiv:1612.08752 [hep-th]}}.
%%CITATION = ARXIV:1612.08752;%%.

\bibitem{Zinoviev:2018juc}
Y.~M. Zinoviev, ``{On massive super(bi)gravity in the constructive approach},''
  \href{http://dx.doi.org/10.1088/1361-6382/aad1fb}{{\em Class. Quant. Grav.}
  {\bfseries 35} no.~17, (2018) 175006},
\href{http://arxiv.org/abs/1805.01650}{{\ttfamily arXiv:1805.01650 [hep-th]}}.
%%CITATION = ARXIV:1805.01650;%%.

\bibitem{Garcia-Saenz:2018wnw}
S.~Garcia-Saenz, K.~Hinterbichler, and R.~A. Rosen, ``{Supersymmetric Partially
  Massless Fields and Non-Unitary Superconformal Representations},''
  \href{http://dx.doi.org/10.1007/JHEP11(2018)166}{{\em JHEP} {\bfseries 11}
  (2018) 166},
\href{http://arxiv.org/abs/1810.01881}{{\ttfamily arXiv:1810.01881 [hep-th]}}.
%%CITATION = ARXIV:1810.01881;%%.

\bibitem{Bonifacio:2016blz}
J.~Bonifacio and K.~Hinterbichler, ``{Kaluza-Klein reduction of massive and
  partially massless spin-2 fields},''
  \href{http://dx.doi.org/10.1103/PhysRevD.95.024023}{{\em Phys. Rev.}
  {\bfseries D95} no.~2, (2017) 024023},
\href{http://arxiv.org/abs/1611.00362}{{\ttfamily arXiv:1611.00362 [hep-th]}}.
%%CITATION = ARXIV:1611.00362;%%.

\bibitem{Sagnotti:2010at}
A.~Sagnotti and M.~Taronna, ``{String Lessons for Higher-Spin Interactions},''
  \href{http://dx.doi.org/10.1016/j.nuclphysb.2010.08.019}{{\em Nucl. Phys.}
  {\bfseries B842} (2011) 299--361},
\href{http://arxiv.org/abs/1006.5242}{{\ttfamily arXiv:1006.5242 [hep-th]}}.
%%CITATION = ARXIV:1006.5242;%%.

\bibitem{deRham:2016plk}
C.~de~Rham, A.~J. Tolley, and S.-Y. Zhou, ``{The $\Lambda_{2}$ limit of massive
  gravity},'' \href{http://dx.doi.org/10.1007/JHEP04(2016)188}{{\em JHEP}
  {\bfseries 04} (2016) 188},
\href{http://arxiv.org/abs/1602.03721}{{\ttfamily arXiv:1602.03721 [hep-th]}}.
%%CITATION = ARXIV:1602.03721;%%.

\bibitem{Karch:2000ct}
A.~Karch and L.~Randall, ``{Locally localized gravity},''
  \href{http://dx.doi.org/10.1088/1126-6708/2001/05/008}{{\em JHEP} {\bfseries
  05} (2001) 008}, \href{http://arxiv.org/abs/hep-th/0011156}{{\ttfamily
  arXiv:hep-th/0011156 [hep-th]}}.
[,140(2000)].
%%CITATION = HEP-TH/0011156;%%.

\bibitem{Karch:2001cw}
A.~Karch and L.~Randall, ``{Localized gravity in string theory},''
  \href{http://dx.doi.org/10.1103/PhysRevLett.87.061601}{{\em Phys. Rev. Lett.}
  {\bfseries 87} (2001) 061601},
\href{http://arxiv.org/abs/hep-th/0105108}{{\ttfamily arXiv:hep-th/0105108
  [hep-th]}}.
%%CITATION = HEP-TH/0105108;%%.

\bibitem{Porrati:2001gx}
M.~Porrati, ``{Mass and gauge invariance 4. Holography for the Karch-Randall
  model},'' \href{http://dx.doi.org/10.1103/PhysRevD.65.044015}{{\em Phys.
  Rev.} {\bfseries D65} (2002) 044015},
\href{http://arxiv.org/abs/hep-th/0109017}{{\ttfamily arXiv:hep-th/0109017
  [hep-th]}}.
%%CITATION = HEP-TH/0109017;%%.

\bibitem{Porrati:2001db}
M.~Porrati, ``{Higgs phenomenon for 4-D gravity in anti-de Sitter space},''
  \href{http://dx.doi.org/10.1088/1126-6708/2002/04/058}{{\em JHEP} {\bfseries
  04} (2002) 058},
\href{http://arxiv.org/abs/hep-th/0112166}{{\ttfamily arXiv:hep-th/0112166
  [hep-th]}}.
%%CITATION = HEP-TH/0112166;%%.

\bibitem{Porrati:2003sa}
M.~Porrati, ``{Higgs phenomenon for the graviton in ADS space},''
  \href{http://dx.doi.org/10.1142/S0217732303011745}{{\em Mod. Phys. Lett.}
  {\bfseries A18} (2003) 1793--1802},
\href{http://arxiv.org/abs/hep-th/0306253}{{\ttfamily arXiv:hep-th/0306253
  [hep-th]}}.
%%CITATION = HEP-TH/0306253;%%.

\bibitem{Aharony:2003qf}
O.~Aharony, O.~DeWolfe, D.~Z. Freedman, and A.~Karch, ``{Defect conformal field
  theory and locally localized gravity},''
  \href{http://dx.doi.org/10.1088/1126-6708/2003/07/030}{{\em JHEP} {\bfseries
  07} (2003) 030},
\href{http://arxiv.org/abs/hep-th/0303249}{{\ttfamily arXiv:hep-th/0303249
  [hep-th]}}.
%%CITATION = HEP-TH/0303249;%%.

\bibitem{Duff:2004wh}
M.~J. Duff, J.~T. Liu, and H.~Sati, ``{Complementarity of the Maldacena and
  Karch-Randall pictures},''
  \href{http://dx.doi.org/10.1103/PhysRevD.69.085012}{{\em Phys. Rev.}
  {\bfseries D69} (2004) 085012},
\href{http://arxiv.org/abs/hep-th/0207003}{{\ttfamily arXiv:hep-th/0207003
  [hep-th]}}.
%%CITATION = HEP-TH/0207003;%%.

\bibitem{Kiritsis:2006hy}
E.~Kiritsis, ``{Product CFTs, gravitational cloning, massive gravitons and the
  space of gravitational duals},''
  \href{http://dx.doi.org/10.1088/1126-6708/2006/11/049}{{\em JHEP} {\bfseries
  11} (2006) 049},
\href{http://arxiv.org/abs/hep-th/0608088}{{\ttfamily arXiv:hep-th/0608088
  [hep-th]}}.
%%CITATION = HEP-TH/0608088;%%.

\bibitem{Aharony:2006hz}
O.~Aharony, A.~B. Clark, and A.~Karch, ``{The CFT/AdS correspondence, massive
  gravitons and a connectivity index conjecture},''
  \href{http://dx.doi.org/10.1103/PhysRevD.74.086006}{{\em Phys. Rev.}
  {\bfseries D74} (2006) 086006},
\href{http://arxiv.org/abs/hep-th/0608089}{{\ttfamily arXiv:hep-th/0608089
  [hep-th]}}.
%%CITATION = HEP-TH/0608089;%%.

\bibitem{deRham:2006pe}
C.~de~Rham and A.~J. Tolley, ``{Mimicking Lambda with a spin-two ghost
  condensate},'' \href{http://dx.doi.org/10.1088/1475-7516/2006/07/004}{{\em
  JCAP} {\bfseries 0607} (2006) 004},
\href{http://arxiv.org/abs/hep-th/0605122}{{\ttfamily arXiv:hep-th/0605122
  [hep-th]}}.
%%CITATION = HEP-TH/0605122;%%.

\bibitem{Kiritsis:2008at}
E.~Kiritsis and V.~Niarchos, ``{Interacting String Multi-verses and Holographic
  Instabilities of Massive Gravity},''
  \href{http://dx.doi.org/10.1016/j.nuclphysb.2008.12.010}{{\em Nucl. Phys.}
  {\bfseries B812} (2009) 488--524},
\href{http://arxiv.org/abs/0808.3410}{{\ttfamily arXiv:0808.3410 [hep-th]}}.
%%CITATION = ARXIV:0808.3410;%%.

\bibitem{Apolo:2012gg}
L.~Apolo and M.~Porrati, ``{On AdS/CFT without Massless Gravitons},''
  \href{http://dx.doi.org/10.1016/j.physletb.2012.07.001}{{\em Phys. Lett.}
  {\bfseries B714} (2012) 309--311},
\href{http://arxiv.org/abs/1205.4956}{{\ttfamily arXiv:1205.4956 [hep-th]}}.
%%CITATION = ARXIV:1205.4956;%%.

\bibitem{Gabadadze:2014rwa}
G.~Gabadadze, ``{The Big Constant Out, The Small Constant In},''
  \href{http://dx.doi.org/10.1016/j.physletb.2014.10.064}{{\em Phys. Lett.}
  {\bfseries B739} (2014) 263--268},
\href{http://arxiv.org/abs/1406.6701}{{\ttfamily arXiv:1406.6701 [hep-th]}}.
%%CITATION = ARXIV:1406.6701;%%.

\bibitem{Gabadadze:2015goa}
G.~Gabadadze and S.~Yu, ``{Metamorphosis of the Cosmological Constant and 5D
  Origin of the Fiducial Metric},''
  \href{http://dx.doi.org/10.1103/PhysRevD.94.104059}{{\em Phys. Rev.}
  {\bfseries D94} no.~10, (2016) 104059},
\href{http://arxiv.org/abs/1510.07943}{{\ttfamily arXiv:1510.07943 [hep-th]}}.
%%CITATION = ARXIV:1510.07943;%%.

\bibitem{Domokos:2015xka}
S.~K. Domokos and G.~Gabadadze, ``{Unparticles as the Holographic Dual of
  Gapped AdS Gravity},''
  \href{http://dx.doi.org/10.1103/PhysRevD.92.126011}{{\em Phys. Rev.}
  {\bfseries D92} (2015) 126011},
\href{http://arxiv.org/abs/1509.03285}{{\ttfamily arXiv:1509.03285 [hep-th]}}.
%%CITATION = ARXIV:1509.03285;%%.

\bibitem{Gabadadze:2017jom}
G.~Gabadadze, ``{Scale-up of $\Lambda_3$: Massive gravity with a higher strong
  interaction scale},''
  \href{http://dx.doi.org/10.1103/PhysRevD.96.084018}{{\em Phys. Rev.}
  {\bfseries D96} no.~8, (2017) 084018},
\href{http://arxiv.org/abs/1707.01739}{{\ttfamily arXiv:1707.01739 [hep-th]}}.
%%CITATION = ARXIV:1707.01739;%%.

\bibitem{Bachas:2017rch}
C.~Bachas and I.~Lavdas, ``{Quantum Gates to other Universes},''
  \href{http://dx.doi.org/10.1002/prop.201700096}{{\em Fortsch. Phys.}
  {\bfseries 66} no.~2, (2018) 1700096},
\href{http://arxiv.org/abs/1711.11372}{{\ttfamily arXiv:1711.11372 [hep-th]}}.
%%CITATION = ARXIV:1711.11372;%%.

\bibitem{Bachas:2018zmb}
C.~Bachas and I.~Lavdas, ``{Massive Anti-de Sitter Gravity from String
  Theory},'' \href{http://dx.doi.org/10.1007/JHEP11(2018)003}{{\em JHEP}
  {\bfseries 11} (2018) 003},
\href{http://arxiv.org/abs/1807.00591}{{\ttfamily arXiv:1807.00591 [hep-th]}}.
%%CITATION = ARXIV:1807.00591;%%.

\bibitem{deRham:2018svs}
C.~De~Rham, K.~Hinterbichler, and L.~A. Johnson, ``{On the (A)dS Decoupling
  Limits of Massive Gravity},''
  \href{http://dx.doi.org/10.1007/JHEP09(2018)154}{{\em JHEP} {\bfseries 09}
  (2018) 154},
\href{http://arxiv.org/abs/1807.08754}{{\ttfamily arXiv:1807.08754 [hep-th]}}.
%%CITATION = ARXIV:1807.08754;%%.

\bibitem{Kotanski:1965zz}
A.~Kotanski, ``{Diagonalization of Helicity-Crossing Matrices},'' {\em Acta
  Physica Polonica} {\bfseries 29} (1966) .

\bibitem{Kotanski:1970}
A.~Kotanski, ``Transversity amplitudes and their application to the study of
  collision of particles with spin,'' {\em Acta Phys. Pol. B1: 45-58(1970)} (1,
  1970) .

\bibitem{Costa:2011mg}
M.~S. Costa, J.~Penedones, D.~Poland, and S.~Rychkov, ``{Spinning Conformal
  Correlators},'' \href{http://dx.doi.org/10.1007/JHEP11(2011)071}{{\em JHEP}
  {\bfseries 11} (2011) 071},
\href{http://arxiv.org/abs/1107.3554}{{\ttfamily arXiv:1107.3554 [hep-th]}}.
%%CITATION = ARXIV:1107.3554;%%.

\bibitem{Kravchuk:2016qvl}
P.~Kravchuk and D.~Simmons-Duffin, ``{Counting Conformal Correlators},''
  \href{http://dx.doi.org/10.1007/JHEP02(2018)096}{{\em JHEP} {\bfseries 02}
  (2018) 096},
\href{http://arxiv.org/abs/1612.08987}{{\ttfamily arXiv:1612.08987 [hep-th]}}.
%%CITATION = ARXIV:1612.08987;%%.

\bibitem{Henning:2017fpj}
B.~Henning, X.~Lu, T.~Melia, and H.~Murayama, ``{Operator bases, $S$-matrices,
  and their partition functions},''
  \href{http://dx.doi.org/10.1007/JHEP10(2017)199}{{\em JHEP} {\bfseries 10}
  (2017) 199},
\href{http://arxiv.org/abs/1706.08520}{{\ttfamily arXiv:1706.08520 [hep-th]}}.
%%CITATION = ARXIV:1706.08520;%%.

\bibitem{Cohen-Tannoudji:1968lnm}
G.~Cohen-Tannoudji, A.~Morel, and H.~Navelet, ``{Kinematical singularities,
  crossing matrix and kinematical constraints for two-body helicity
  amplitudes},''
\href{http://dx.doi.org/10.1016/0003-4916(68)90243-1}{{\em Annals Phys.}
  {\bfseries 46} no.~2, (1968) 239--316}.
%%CITATION = APNYA,46,239;%%.

\bibitem{Kotanski1968}
A.~Kota{\'{n}}ski, ``Kinematical singularities of the transversity
  amplitudes,'' \href{http://dx.doi.org/10.1007/BF02819831}{{\em Il Nuovo
  Cimento A (1971-1996)} {\bfseries 56} no.~3, (Aug, 1968) 737--754}.

\bibitem{Lee:1977eg}
B.~W. Lee, C.~Quigg, and H.~B. Thacker, ``{Weak Interactions at Very
  High-Energies: The Role of the Higgs Boson Mass},''
\href{http://dx.doi.org/10.1103/PhysRevD.16.1519}{{\em Phys. Rev.} {\bfseries
  D16} (1977) 1519}.
%%CITATION = PHRVA,D16,1519;%%.

\end{thebibliography}\endgroup

\end{document}